\documentclass[pra,aps,superscriptaddress,twocolumn,letter,nopacs,longbibliography,notitlepage]{revtex4-1}
\usepackage{graphicx,color,amsmath,amsfonts,enumerate,amsthm,amssymb,mathtools,enumitem,thmtools,hyperref,mathdots,enumitem,centernot,bm,soul,bbm,stmaryrd}

\usepackage[capitalise, noabbrev]{cleveref}

\usepackage{multirow}
\usepackage{amsmath}
\usepackage{times}
\usepackage{tikz,pgfplots}
\pgfplotsset{compat=1.17}
\usetikzlibrary{quantikz}
\hypersetup{colorlinks=true,linkcolor=teal,citecolor=teal,urlcolor=teal}
\usepackage{bbm}
\usepackage{algorithm,algpseudocode}
\usepackage{algorithmicx}
\usepackage[skins]{tcolorbox}

\usepackage[normalem]{ulem}

\theoremstyle{plain}

\theoremstyle{definition}

\newcommand{\commentblock}[1]{}

\newcommand{\ud}{{\textrm{d}}}

\definecolor{augustine}{rgb}{0,.5,.5}

\newcommand{\newtext}[1]{{\color{black}{#1}}}

\begin{document}

\title{
Local integrals of motion and the stability of many-body localisation in Wannier-Stark potentials}


\author{C.~Bertoni}
\affiliation{Dahlem Center for Complex Quantum Systems, Freie Universit\"{a}t Berlin, Germany}
\author{J.~Eisert}
\affiliation{Dahlem Center for Complex Quantum Systems, Freie Universit\"{a}t Berlin, Germany}
\address{Helmholtz-Zentrum Berlin f{\"u}r Materialien und Energie, 14109 Berlin, Germany}
\author{A.~Kshetrimayum}
\affiliation{Dahlem Center for Complex Quantum Systems, Freie Universit\"{a}t Berlin, Germany}
\address{Helmholtz-Zentrum Berlin f{\"u}r Materialien und Energie, 14109 Berlin, Germany}
\address{Theory Division, Saha Institute of Nuclear Physics, 1/AF Bidhannagar, Kolkata 700 064, India}
\author{A.~Nietner}
\affiliation{Dahlem Center for Complex Quantum Systems, Freie Universit\"{a}t Berlin, Germany}
\address{Helmholtz-Zentrum Berlin f{\"u}r Materialien und Energie, 14109 Berlin, Germany}
\author{S.~J.~Thomson}
\affiliation{Dahlem Center for Complex Quantum Systems, Freie Universit\"{a}t Berlin, Germany}
\address{Current address: IBM Research - UK, Winchester, United Kingdom, SO21 2JN}

\date{\today}



\begin{abstract}

Many-body localisation in disordered 
systems in one spatial dimension is typically understood in terms of the existence of an extensive number of (quasi)-local integrals of motion (LIOMs) which are thought to decay exponentially with distance and interact only weakly with one another. By contrast, little is known about the form of the integrals of motion in disorder-free systems which exhibit localisation. Here, we explicitly compute the LIOMs for disorder-free localised systems, focusing on the case of a linearly increasing potential. We show that while in the absence of interactions, the LIOMs decay faster than exponentially, the addition of interactions leads to the formation of a slow-decaying plateau at short distances. We study how the localisation properties of the LIOMs depend on the linear slope, finding that there is a significant finite-size dependence, and present evidence that adding a weak harmonic potential does not result in typical many-body localisation phenomenology. By contrast, the addition of disorder has a qualitatively different effect, dramatically modifying the properties of the LIOMS. 
\end{abstract}
\maketitle


\section{Introduction}

It is by now well-established that the presence of disorder in a low-dimensional quantum mechanical system can lead to localisation. Such notions of localisation are accompanied 
by the absence of thermalization via
Hamiltonian out-of-equilibrium dynamics
\cite{PolkovnikovReview,christian_review,AbaninEtAlRMP19,1408.5148}. Disorder-induced localisation was first discovered in non-interacting systems~\cite{Anderson58}, 
in a phenomenon now known as 
\emph{Anderson localisation}. 
In recent years, this framework has been extended to include systems with many-body interactions, leading to the celebrated phenomenon of 
\emph{many-body localisation} (MBL)~\cite{Fleishman+80,Basko+06,Huse+13,HuseNandkishoreOganesyanPRB14,Altman+15,Luitz+15,Alet+18,AbaninEtAlRMP19,1409.1252}. 

Many-body localised systems exhibit a variety of characteristic 
and unusual properties, such as a logarithmic growth of entanglement entropy with time~\cite{Bardarson+12,Prosen_localisation,SerbynPapicAbaninPRL13} or of other correlation measures~\cite{Accessible},
or a persistent density imbalance, a phenomenon that  has been observed in experiments~\cite{Schreiber+15}. 

While there is strong evidence that many-body localisation is a stable phase of matter~\cite{Imbrie16a}, recently the claim that such localisation is truly stable in the thermodynamic limit has been called into question~\cite{Suntajs+20a,Suntajs+20b,Sels+21a,Sels+21b,Sierant+21,Long+22,KieferEmmanouilidis+22,Sierant+22,BeraMBLinstability2023}. This is a question
that will ultimately be difficult to decide;
it has been suggested that establishing any genuine properties of the MBL phase transition is likely to be extremely challenging for current numerical techniques~\cite{Panda+20}.

The study of so-called \emph{Wannier-Stark localisation} in translationally invariant free-electron systems has a long history~\cite{Wannier62} and has recently been investigated in solid state physics~\cite{Schmidt+18,Guo+21,Berghoff+21}. In the context of many-body systems, the question of whether \emph{disorder-free} many-body systems can exhibit genuine localisation phenomena was already studied as far back as 2015~\cite{Papic+15}, but came to prominence with a pair of simultaneous works published in 2019~\cite{vanNieuwenburg+19,Schulz+19}.
Ref.~\cite{vanNieuwenburg+19} has shown that a disorder-free system with a linear potential exhibits a localised phase which is 
smoothly connected to the well-studied conventional MBL phase in a disordered model, while Ref.~\cite{Schulz+19}
has demonstrated that while the disorder-free system with a purely linear potential did not exhibit typical MBL properties, the additional of a weak quadratic confining potential leads to an entanglement growth which is consistent with conventional MBL. These 
observations have sparked widespread interest in localisation and related phenomena such as time crystals in disorder-free systems~\cite{MonroeNaturestarkMBL,Gunawardana2021,KshetrimayumstarkTC,PizziPRB2020,Wybo+21}. 
\newtext{The proposed physical mechanism behind the phenomenology of many-body localisation is the appearance of a form of emergent integrability. An integrable system is one with an extensive number of local conserved quantities, such that the initial condition may be inferred at any time by local measurements. Many-body localisation is then described by the emergence of an extensive set of quasi-local conserved quantities, called (quasi-) \emph{local integrals of motion} (LIOMs). These are conserved quantities which are local up to tails decaying exponentially in real space. Much of the phenomenology of MBL can be recovered analytically by assuming such quantities exist \cite{Imbrie+17} and studies of LIOMs in disordered and quasiperiodic systems have so far verified their emergence from macroscopic models and validated their use in the mathematical description of MBL \cite{Rademaker+16, Rademaker+17,Accessible, Thomson+18,Thomson+20, Thomson+21}, and experimental probes of their properties have been put forward \cite{lu2023measuring}. For these reasons, explicitly constructing the LIOMs leads to valuable insight into the physics of MBL which can otherwise only be gleaned by individual probes such as slow growth of entanglement and particle imbalance. Moreover, the mathematical description of LIOMs is independent of the specifics of the Hamiltonian, notably, whether it is disordered or not, and it is not immediately clear why disorder would be necessary for their emergence. Their study in both disordered and non-disordered can then better elucidate the role that disorder plays in the phenomenology of MBL. }

Motivated by this growing interest, in this work, we examine the localisation properties of disorder-free systems in one dimension through the lens of their local integrals of motion. We will examine the localisation properties of these LIOMs using several different measures. We note for completeness that disorder-free localisation has recently been studied in the context of \emph{lattice gauge theories}~\cite{Smith+17,Brenes+18,Karpov+21,Halimeh+22a,Halimeh+22b,Lang+22,Chakraborty+22}, where the conserved charges of the theory give rise to an effective disorder, 
however, this is not the form of disorder-free localisation which we consider in this work. We also wish to emphasise at the outset that this work studies the properties of individual LIOMs, in contrast to other works on 
many-body localisation which typically focus on non-equilibrium dynamics. The relaxation dynamics of an arbitrary far-from-equilibrium state is governed not only by the spatial extent of the LIOMs, but also their overlap with one another, a crucial point which we shall return to later. 

The structure of this work is as follows. In Section~\ref{sec.model}, we will introduce the model that we will be working with, and the different choices of potential that we study in this work. In Section~\ref{sec.method}, we will detail how the LIOMs are computed using the \emph{tensor flow equation} (TFE) method~\cite{Thomson+21}; they introduce an alternative measure to probe how quasi-local the unitary transform used to diagonalise the Hamiltonian really is. In Section~\ref{sec.results}, we will present numerical results obtained using the TFE method, before concluding in Section~\ref{sec.conclusion} by discussing our results in the context of other contemporary work.

\section{Model}
\label{sec.model}
\newtext{In this section, we present the models which we examine in this work, namely XXZ models with an added clean potential which replaces the more standard disordered field in traditional models of MBL. In the last subsection, we discuss properties of these potentials which sets them apart from the disordered ones.}
We start from the Hamiltonian
\begin{align}\label{eq: model}
\mathcal{H} = \sum_i h_i n_i + J \sum_{i} (c^{\dagger}_i c_{i+1} + H.c.) + \Delta_0 \sum_{i} n_i n_{i+1},
\end{align}
which describes a one-dimensional chain of spinless fermions with open boundary conditions,
where $h_i$ is the varying on-site potential, $J$ is the nearest-neighbour hopping and $\Delta_0$ is the nearest-neighbour interaction strength. In all of the following, we shall set $J=1$ as the unit of energy. This model is equivalent to the XXZ spin chain via a Jordan-Wigner transform, however, fermionic algebra is much more suitable\footnote{This is because the method involves computing nested commutators, and the (anti)commutator of two fermions is simply a number, while the commutator of two spin operators is another operator. Fermionic algebra therefore dramatically reduces the number of terms that arise during the calculation, simplifying it considerably.} for the method we will introduce in Section~\ref{sec.method}. We will examine three different choices of potential, detailed below.

\subsection{Linear potential}

The first potential we will look at is the purely linear potential, where $h_i = F i$ and $F$ measures the tilt of the lattice. 
We first review the known behaviour of the model with $\Delta_0 = 0$. As detailed in Ref.~\cite{vanNieuwenburg+19}, this model is Wannier-Stark localised for all $F \neq 0$, and can be diagonalised by the transform
\begin{align}
b_m = \sum_j \mathcal{J}_{j-m}(2J/F) c_j,
\end{align}
where $\mathcal{J}_n$ is a Bessel function of the first kind. This leads to a diagonal Hamiltonian
\begin{align}
\tilde{\mathcal{H}} = - \sum_m F m b^{\dagger}_m b_m .
\end{align}
As the Bessel functions $x\mapsto\mathcal{J}_n(x)$ decay faster than exponential ($|\mathcal{J}_n| < \textrm{e}^{-|n|}$ for $x \ll n$), this non-interacting system is Wannier-Stark localised for all values of $F \neq 0$. It is interesting to note that this mechanism is quite different from the case of Anderson localisation, and by means of a gauge transform the linear potential can be mapped onto a time-dependent vector potential rather than an on-site field.

\begin{figure}[t!]
    \centering
    \includegraphics[width=\linewidth]{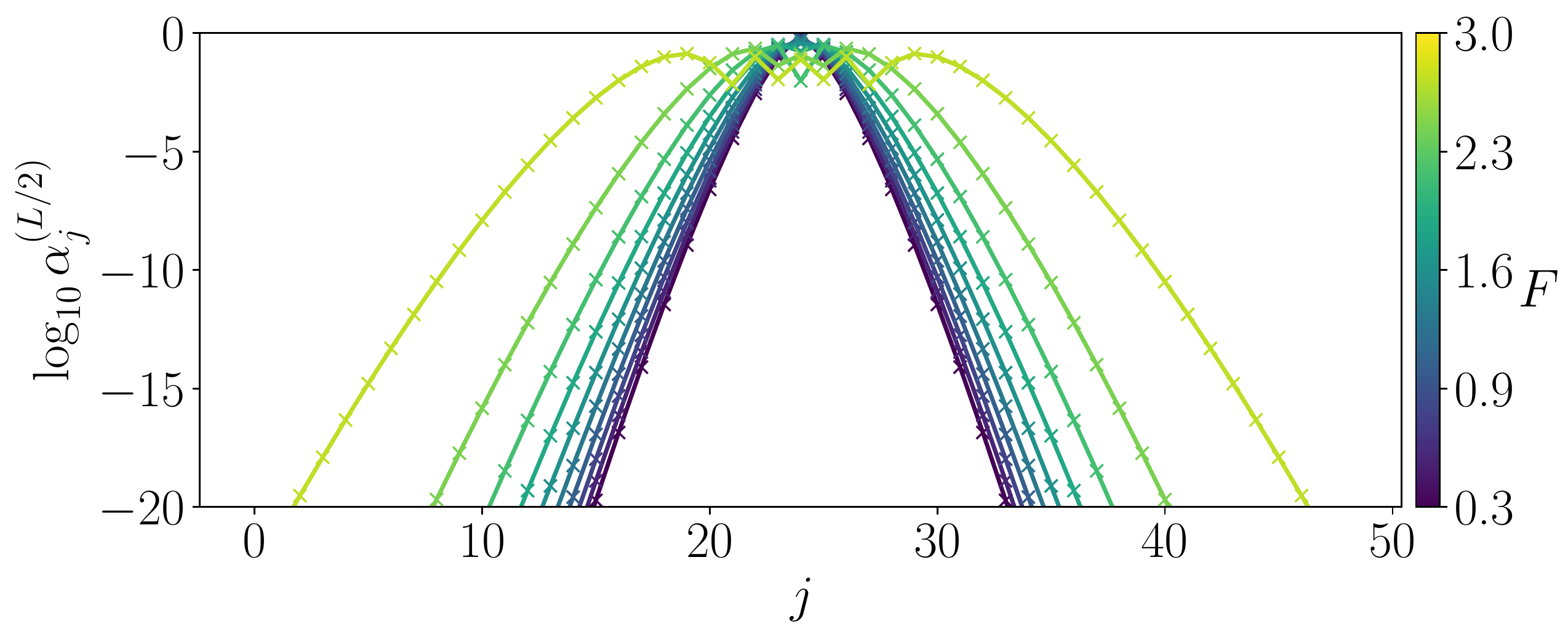}
    \caption{The real-space decay of a local integral of motion centred in the middle of a one-dimensional non-interacting system of length $L=64$ with a linearly increasing potential of slope $F$, and hopping $J=1$. The integrals of motion are given by Bessel functions of the first kind, as shown in Eq.~(\ref{eq.bessel}), resulting in a faster-than-exponential decay in real space and a highly non-trivial oscillatory structure close to the centre of the $l$-bit.}
    \label{fig.bessel}
\end{figure}

The (local) integrals of motion of this non-interacting system are given by $\tilde{n}_i = b^{\dagger}_i b_i$ and can be expressed in the microscopic basis using the transform specified above
\begin{equation}
    \begin{aligned}
    \tilde{n}_i &= b^{\dagger}_i b_i = \sum_{j,k} \mathcal{J}_{j-i} (2J/F) \mathcal{J}_{k-i}(2J/F) c^{\dagger}_j c_k,  \\
    & =: \sum_{j} \alpha^{(i)}_j c^{\dagger}_j c _j + \sum_{j \neq k} \beta^{(i)}_{j,k} c^{\dagger}_j c_k,
    \label{eq.bessel}
\end{aligned}
\end{equation}
where we denote the diagonal terms $\alpha^{(i)}_j$ and the off-diagonal terms $\beta^{(i)}_{j,k}$, a notation which we shall use throughout. The diagonal terms for a local integral of motion prepared on the central site of a one-dimensional chain are shown in Fig.~\ref{fig.bessel}, where it is clearly visible that they decay with distance in a faster-than-exponential manner. It is also interesting to note that close to the centre site where the $l$-bit is located, the diagonal coefficients $\alpha^{(i)}_j$ exhibit an unusual oscillatory behaviour that even in a small system enables them to be clearly distinguished from the exponentially localised $l$-bits expected in an Anderson localised system. It is likely that this form should be experimentally measurable using techniques that have already been applied to investigate Anderson localisation in ultra-cold atomic gases~\cite{Billy+08}, although to date this has not (to our knowledge) been attempted.

For non-zero values of $\Delta_0$, this model was first studied in Refs.~\cite{vanNieuwenburg+19,Schulz+19} and found to remain localised above a critical field strength $F_c$, albeit with a faster than logarithmic growth of the entanglement entropy due to the existence of a large number of many-body degeneracies in the energy spectrum which stem from the perfect translational invariance of this model. To obtain more conventional MBL-like growth of the entanglement entropy in time, it is necessary to lift these degeneracies, either through the addition of weak randomness or by adding some curvature to the potential. 

\subsection{Curved potential}

In order to lift the many-body degeneracies which occur in a system with a purely linear potential, we follow an approach similar to Ref.~\cite{Schulz+19} and modify the on-site potential to include some weak curvature of the form 
\begin{align}
    h_i = Fi + \alpha \left( \frac{i}{L} \right)^2,
\end{align}
where $L$ is the system size and $\alpha$ represents the strength of the curvature. 
For small values of $\alpha$, the system breaks translation invariance only weakly, and so the main effect of the curvature is to lift the degeneracies and stabilise localisation, as we briefly demonstrate in the Appendix, section II. Note that due to the factor of $1/L^2$, this curvature vanishes in the thermodynamic limit. This is required in order to prevent the curvature dominating over the linear slope for large systems, but has the effect of making it difficult to extract a well-defined behaviour in the thermodynamic limit (insofar as this limit exists at all for systems with linearly increasing potentials). An alternative form of curved potential without this factor, more akin to the harmonic confining potentials used in ultracold atomic gas experiments, is briefly studied in the Appendix, section III for completeness.
It has recently been argued that conventional many-body localisation in a random potential in two dimensions can be stabilised by superimposing a weakly curved potential~\cite{Foo+22}, due to the curvature making it difficult for long-range resonances to arise and limiting their effectiveness at delocalising the system. It is therefore of significant interest to understand the role of weakly curved 
potentials in potentially stabilising localisation.

\subsection{Weakly disordered potential}

We also consider a linear potential with (weak) disorder superimposed on top of it, in order to investigate the effect of randomness on Wannier-Stark localisation and search for a crossover to conventional MBL-like behaviour, i.e., quantities which are exponentially localised, in contrast to the Bessel-function-like localisation of Eq.~(\ref{eq.bessel}). We will use a 
potential of the form
\begin{align}
    h_i = F i + \varepsilon_i,
\end{align}
with the randomly chosen variable $\varepsilon_i \in [-d,d]$. A potential of this form has been considered in Ref.~\cite{vanNieuwenburg+19}, who have shown 
that the Wannier-Stark many-body localised phase appears to be smoothly connected to the conventional disordered MBL phase, however, 
there is as yet no understanding of how the LIOMs behave in the presence of both a linear potential and additional disorder.
The addition of disorder is highly non-trivial, and will not necessarily strengthen localisation (as it may potentially introduce rare resonant regions), and will lead to a competition between Bessel-function-like Wannier-Stark localisation and exponentially-decaying (many-body) Anderson localisation.

\subsection{The role of degeneracies}
\label{sec.degen}

In selecting the potentials in this section, we have only briefly mentioned the role of degeneracies in the spectrum. Let us now elaborate on their importance.
An interesting phenomenon that has emerged from previous research on 
many-body localisation is the crucial role that degenerate states play in the localisation of the system. While a simple linear slope exhibits some signs of localisation, such as persistent imbalance and Poisson level statistics~\cite{vanNieuwenburg+19}, other signs of localisation are absent. Notably, the entanglement entropy quickly saturates as opposed to showing the slow logarithmic growth typical of disordered many-body localised systems~\cite{Schulz+19, RYao21}. A natural question to ask is then whether adding curvature to a linearly growing potential is sufficient to fully replicate the physics of a disordered potential. 
To better explain how a non-degenerate potential may lead to slower thermalization, consider a local spin Hamiltonian on $n$ sites of the form
\begin{equation}
    H(\gamma)=D+\gamma V
\end{equation}
where $D$ is diagonal in the computational basis and $\|V\|=\|D\|=1$ and $\gamma<1$. Here, the off diagonal part $V$ is treated as a perturbation of the diagonal part $D$. In the case of the model in \cref{eq: model}, the first term corresponds to $D$ and the other two to $V$, and after suitable normalizations we have $1/\gamma\sim \max_i h_i$. Clearly, $D$ is trivially ``localized", as the local integrals of motion are given by the single qubit Pauli operators $\sigma_z^i$ one each site. We now want to argue that if $D$ has no degeneracies, adding a small off diagonal local perturbation $\gamma V$ can only affect this weakly. Consider the operator
\begin{equation}\label{eq:SW_generator}
    A_{k,l}=i \frac{V_{k,l}}{D_k-D_l} 
\end{equation}
where $k,l$ are computational basis states and $D_k=D_{k,k}$. One may verify that
\begin{equation}
    e^{i\gamma A}H e^{-i\gamma A}=D+i\gamma^2[A,V]+O(\gamma^3)
\end{equation}
where by $O(\gamma^3)$ we mean terms the norm of which is bounded above by $O(\gamma^3)$. Assuming that the denominator in \cref{eq:SW_generator} does not vanish (that is, there are no degenerate states in $D$ coupled by $V$), as long as $V$ and $D$ are local, so is $A$, by standard Lieb-Robinson bound type arguments \cite{Lieb_1972}, this implies that the dressed Pauli operators $\tau_i=e^{-i\gamma A}\sigma_z^i e^{i\gamma A}$ are quasilocal, and furthermore $\|[H,\tau_i]\|\sim \gamma^3$.
This approach, followed by an iteration to higher orders, has been used in several works relating to many-body localization in the case of disordered potentials \cite{De_Roeck_2014,Imbrie+17, Thiery17} and to show slow thermalisation in various regimes \cite{Abanin_2017}.
The linear potential has no degeneracies between states separated by a single hop, as such states have an energy difference of at least $F$. These states are the only ones directly coupled by the off-diagonal part $V$ (in this case, the XXZ Hamiltonian) at first order in perturbation theory. This means that we will have $\delta\propto 1/F$, as the denominator in \cref{eq:SW_generator} is proportional to $F$. Then we have that for $F$ large enough, there are quasi-local operators $\tilde{\sigma}_z^i$ such that $[\tilde{\sigma}_z^i, H]\sim {1}/{F^2}$. These are then quasi-local approximately conserved quantities that should hinder thermalisation. 

Notice how the absence of degeneracies or \emph{resonances} (that is, nearly degenerate states) in $D$ for states connected by $V$ is essential to control the norm of $A$. If we wished to iterate this procedure to remove the off diagonal potential to higher orders, we would now need states connected by the new potential $[A,V]$ to be non degenerate, but if $V$ is nearest-neighbour, it is easy to see that $[A,V]$ has next to nearest neighbour terms. We would then need states connected by flipping spins on larger and larger portions of the lattice to be non degenerate. Nevertheless, states separated by multiple hops in the linear potential can be degenerate. For example, the state vectors $|0,1,0,1,0,1,0\rangle$ and  $|1,0,0,1,0,0,1\rangle$ are separated by two hops and have exactly the same energy, hence they cannot be eliminated in second order perturbation theory. As a matter of fact, the number of degenerate states must scale exponentially in the system size, since the number of possible many-body eigenvalues the potential can give rise to is only linear in the system size, while the number of possible states is exponential.
Contrariwise, in a sufficiently strongly disordered potential the probability of finding two exactly degenerate states is zero, and while the energies can be arbitrarily close to each other with non vanishing probability, these resonant states can be rare enough as not to cause thermalization \cite{Imbrie+17}.
It is then expected that the pure linear potential cannot reproduce the full physics of disordered potentials, as the proliferation of dengeneracies with increasing system size renders it impossible for perturbative approaches to converge. This in turn suggests that the many-body eigenstates should not behave like the (localised) single-particle eigenstates, although this problem may be avoided for small enough systems that the degeneracies are sufficiently few in number. 

The rationale behind adding curvature (cf.~Ref.~\cite{Schulz+19}) is then to remove these higher order exact degeneracies. Many degeneracies are indeed removed, but by the same argument as above we can say that the number of possible values that the curved potential can take is at most quadratic in the system size -- the possible values are of the form $Fa+\alpha/L^2 b$, with $a<L, b<L^2$ -- hence there must still be exponentially many exactly degenerate states. \newtext{In general, for Wannier-Stark like polynomial potential of degree $q$, the diagonal of $D$ can take at most $\sim L^q$ unique parameters. One would then have to take a polynomial of degree $\sim L/\log(L)$ in order to have enough possible unique values. By this argument we can conclude that no finite degree polynomial potential can replicate the physics of a disordered potential. }

By this heuristic, it is reasonable to suggest that while for a given system size the curved potential might display \emph{more} signs of localisation than the purely linear slope -- a symptom of being able to perturbatively remove higher order couplings -- it also cannot reproduce the full physics of disordered potentials for infinite times and system sizes. This argument can be extended \emph{ad infinitum} by adding further non-linear terms in order to lift all degeneracies at a given system size, but ultimately the exponential number of states must always be shared between a polynomial number of possible energy eigenvalues, meaning that for a sufficiently large system there will always be enough degeneracies that perturbative approaches will fail to converge, suggestive of delocalisation. This stands in dramatic contrast with the case of conventional (random) disorder.

\section{Method}
\label{sec.method}
Computing the local integrals of motion is equivalent to diagonalising the Hamiltonian. In this work we perform the diagonalisation using the \emph{tensor flow equation} (TFE) method~\cite{Thomson+21}, based on G\l{}azek-Wegner-Wilson flow methods~\cite{Brockett91,Chu+90,Chu94,Glazek+93,Glazek+94,Kehrein07} as implemented in the \texttt{PyFlow} package~\cite{PyFlow}. In this section we shall briefly summarise the key aspects of the method.
We note that there is no unique prescription for computing the LIOMs of a many-body system, and other complementary methods exist based 
around discrete displacement transforms~\cite{Rademaker+16,Rademaker+17} and time-averaged observables~\cite{Chandran+15,Goihl18,Singh+21}. 

\subsection{Diagonalizing the Hamiltonian}
\label{sec.TFE_diag}
We first separate the Hamiltonian into a diagonal part
\begin{equation}
\begin{aligned}
\mathcal{H}_0 &= \sum_i h_i n_i + \Delta_0 \sum_{i} n_i n_{i+1} ,
\end{aligned}
\end{equation}

and an off-diagonal part
\begin{align}
V &= J \sum_{i} (c^{\dagger}_i c_{i+1} + H.c.).
\end{align}
The Hamiltonian is diagonalised by the  continuous unitary transform
\begin{equation}
    \begin{aligned}
\frac{\ud \mathcal{H}}{\ud l} = [\eta(l), \mathcal{H}(l)].
\label{eq.flow}
\end{aligned}
\end{equation}

where $\eta(l)=[\mathcal{H}_0(l),V(l)]$ is the so-called \emph{Wegner generator}~\cite{Wegner94,Bach}. This form of generator works using the principle of energy scale separation, first eliminating off-diagonal terms which couple significantly different energy sectors of the Hamiltonian before successively eliminating lower and lower energy degrees of freedom, at the cost of generating new terms (both diagonal and off-diagonal) at intermediate flow times, which must be kept track of throughout the procedure. This method has a long history~\cite{Brockett91,Chu+90,Chu94,Glazek+93,Glazek+94,Kehrein07,PhysRevLett.100.130501}
as well as a sound mathematical underpinning and convergence guarantees
~\cite{Bach,Hastings22}.
Various different implementations of the method have been used to study many-body localisation extensively in recent years in both time-independent~\cite{Monthus16,Pekker+17,Thomson+18,Kelly+20,Thomson+20,Thomson+20b,Thomson+21} and time-dependent~\cite{Tomaras+11,Verdeny+13,Vogl+19,Thomson+20c} settings. It is also closely related to the method of discrete displacement transforms~\cite{Rademaker+16,Rademaker+17}. 

In its most recent implementation~\cite{Thomson+21,PyFlow}, the Hamiltonian can be written in tensor form and the commutator in Eq.~(\ref{eq.flow}) 
can be recast in terms of tensor contractions, which can be efficiently computed on modern computer hardware using highly optimised linear algebra libraries, eliminating the need to analytically determine the flow equations, therefore bypassing one of the most cumbersome aspects of flow equation techniques. In this work, we run our simulations on 
\emph{graphics 
processing units} (GPUs), which are highly optimised for massively parallel linear algebra operations. We numerically integrate the flow equation until all off-diagonal terms present in the initial Hamiltonian have decayed to less than a fraction $\varepsilon = 10^{-6}$ of their original value, or the maximum flow time of $l=25$ is reached. We store up to $q=7500$ steps for the smallest system sizes and $q=3500$ steps for the largest system sizes, with the steps spaced logarithmically in flow time. Detailed convergence and error measures are shown in the Appendix, section I.

It is worth emphasising that the principle of separation of energy scales which underpins this method has serious consequences in systems with exact degeneracies. The transform cannot remove off-diagonal couplings which connect degenerate states, and while the linear potential does not have single-particle degeneracies, it \emph{can} have many-body degeneracies, as discussed in Sec.~\ref{sec.degen}. We can thus anticipate that there may be regimes where this transform fails to remove all off-diagonal couplings. The existence of such off-diagonal couplings which connect different degenerate states signifies that fermions are able to move freely throughout the lattice, implying delocalisation. We shall refer to these as `resonant terms'.

For a non-interacting system, this transform can be performed exactly and 
leads to a (quadratic) diagonal Hamiltonian in 
the $l \to \infty$ limit. For an interacting system, this transform will generate successively higher order terms and the final Hamiltonian will be given by
\begin{equation}
\begin{aligned}
\tilde{\mathcal{H}} &= \sum_i \tilde{h}_i \tilde{n}_i + \frac12 \sum_{i,j} \Delta_{i,j} \tilde{n}_i \tilde{n}_j \\&+ \sum_{i,j,k} \xi_{i,j,k} \tilde{n}_i \tilde{n}_j \tilde{n}_k + \dots ,
\label{eq.Hdiag}
\end{aligned}
\end{equation}
where the tilde notation signifies that quantities are written in the diagonal ($l \to \infty$) basis. In general we will neglect all terms of 3rd order and more, i.e., we keep the 2-body $\Delta_{i,j} \tilde{n}_i \tilde{n}_j$ terms but neglect the 3-body and higher terms, in order to keep the computation tractable. The source term which generates the 3-body interaction scales at most like $\sim J \Delta_0^2$~\cite{Thomson+21}, so for weak interactions with $\Delta_0 \ll 1$, the 3-body terms are typically negligible. The exception to this is if the system is delocalised due to the interactions, in which case these neglected higher-order terms become relevant during the flow procedure and the transform may deviate significantly from unitarity, resulting in divergent terms and a loss of accuracy, similar in spirit to a divergent renormalisation group flow towards a strong coupling fixed point. The essential idea is that the unitary transform can add -- as well as remove -- resonant off-diagonal terms during the flow, and if these terms are added faster than they can be removed, the method will fail to converge. The breakdown of the method can then be associated with proximity to a delocalisation transition and the proliferation of resonances. We can already anticipate that for a fixed interaction strength, this method should break down below some critical field strength $F_c$ where the system becomes delocalised. This can be made more precise through a self-consistent error estimation method shown in the Appendix, section I.

\subsection{Obtaining local integrals of motion (LIOMs)}

The Hamiltonian in Eq.~(\ref{eq.Hdiag}) is diagonal 
in terms of the operators $\tilde{n}_i$, which we shall call 
(quasi)-\emph{local integrals of motion} (LIOMs), anticipating that this system will display signatures of localised behaviour. One way we can gain insight as to the nature of the phase is to examine the real-space support of these objects, which we can compute by transforming them from the diagonal ($l \to \infty$) basis back into the initial real-space basis ($l = 0$).
In order to do this, we store the flow of the generator $\eta(l)$ which diagonalises the Hamiltonian, and then we can apply the same series of infinitesimal transforms in reverse to move from the diagonal basis back into the microscopic basis. The flow equation for a LIOM operator $\tilde{n}_i$ is given by
\begin{align}
\frac{\ud \tilde{n}_i}{\ud l} = [\eta(l), \tilde{n}_i(l)],
\end{align}
The end result of this will be the LIOM operator $\tilde{n}_i$ located on site $i$ expressed in terms of the original microscopic fermionic operators $n_j$,
\begin{equation}
\begin{aligned}
\tilde{n}_i &= \sum_j \alpha^{(i)}_j n_j + \sum_{j,k} \beta^{(i)}_{j,k} c^{\dagger}_j c_k + \sum_{j,k} \gamma^{(i)}_{j,k} n_j n_k \\&+ \sum_{j,k,p,q} \zeta^{(i)}_{j,k,p,q} c^{\dagger}_j c_k c^{\dagger}_p c_q + ...
\label{eq.LIOM}
\end{aligned}
\end{equation}
where the superscripts $(i)$ denote that the original LIOM has been centred around lattice site $i$. This has a similar form to Eq.~(\ref{eq.bessel}), but now with additional higher-order terms due to the interactions present in the Hamiltonian. In the absence of interactions, only the $\alpha$ and $\beta$ coefficients are non-zero and the LIOMs can be calculated exactly. On the other hand, for an interacting system this equation does not close and we have to make a suitable truncation. This truncation must be consistent with the truncation made for the Hamiltonian, e.g., if we neglect all terms of a given order in in the flow of the Hamiltonian, we cannot consistently include these terms in the flow of the number operator. 

We will be interested in the behaviour of the coefficients $\alpha^{(i)}_j$, $\beta^{(i)}_{j,k}$, $\gamma^{(i)}_{j,k}$ and $\zeta^{(i)}_{j,k,p,q}$. In the following we will look at their norm and real-space support. In all of the following, we shall compute only a single LIOM located on the central site of the one-dimensional chains that we consider, in order to avoid boundary effects.

\subsection{Operator spreading in flow time}

Another way to get a measure of the locality of the system is to apply the unitary operator which diagonalises the Hamiltonian to an operator which is local in the original basis, e.g., an on-site number operator. We can again do this by applying the unitary transform used to diagonalise the Hamiltonian to the desired operator in order to transform it into the diagonal basis. This procedure does not require storing the full unitary transform, as it can be performed on-the-fly simultaneously with the diagonalisation process. The flow equation for the operator $n_i$ is given by
\begin{align}\label{eq:number_flow}
\frac{\ud n_i}{\ud l} = [\eta(l), n_i(l)].
\end{align}
The end result of this will be an initially local operator $n_i$ located on site $i$ expressed in terms of the LIOMs $\tilde{n}_j$. We denote this operator by $n_i(l \to \infty) = n_i^*$ to signify that it is the microscopic number operator expanded in terms of operators from the diagonal basis, and it takes the form
\begin{equation}
\begin{aligned}
    n_i(l \to {\infty}) &= \sum_j \tilde{\alpha}^{(i)}_j \tilde{n}_j + \sum_{j,k} \tilde{\beta}^{(i)}_{j,k} \tilde{c}^{\dagger}_j \tilde{c}_k \\&+ \sum_{j,k} \tilde{\gamma}^{(i)}_{j,k} \tilde{n}_j \tilde{n}_k + \sum_{j,k,p,q} \tilde{\zeta}^{(i)}_{j,k,p,q} \tilde{c}^{\dagger}_j \tilde{c}_k \tilde{c}^{\dagger}_p 
\tilde{c}_q + \dots\,.
\label{eq.fwd_liom}
\end{aligned}
\end{equation}
This has essentially the same structure as Eq.~(\ref{eq.LIOM}), but the role of the number operator and LIOMs is reversed - note the interchange of the operators $n_i$ and $\tilde{n}_i$. 
This operator, which from now on we will call a \emph{transformed number operator}, is \emph{not} an integral of motion, nevertheless we expect that it shares many of the properties of the LIOMs. 

We shall call these ``transformed number operators'' to distinguish them from the LIOMs considered in the previous section. If the local integrals of motion $\tilde n_i$ are obtained by evolving the number operators backwards using the unitary $U$ which diagonalises the Hamiltonian, i.e., $\tilde n_i=U^{\dagger}n_iU$, then the transformed number operators are obtained by performing the same evolution, but forward, i.e., $Un_iU^{\dagger}$. It is then reasonable to suggest that $n_i^*$ should have similar locality properties as $\tilde n_i$. 
More precisely, a way of encoding the property that $U$ maps local number operators to quasi-local operators -- the LIOMs -- is that $U$ should satisfy a property akin to a Lieb-Robinson bound. That is, given two observables $A$ and $B$ supported on regions separated by a distance $r$, we need to have $||[U^{\dagger}AU, B]||_{\infty}\leq C\,e^{-r/\xi}$, for some characteristic length $\xi$ and a constant $C$. This is a 
comparably weak definition of quasi-locality\footnote{Compare
also the discussion in 
Ref.~\cite{Goihl18}.}, but it should be expected that it holds for many-body localised systems. Then it is true that the inverse evolution also satisfies the same, since 
\begin{equation}
||[UAU^{\dagger}, B]||_{\infty}=||[A, U^{\dagger}BU]||_{\infty}\leq C\,e^{-r/\xi}. 
\end{equation}
This means that, in this definition, $n_i(l\to\infty)$ is quasilocal.
We thus expect that many commonly used locality probes will yield the same results for $n_i^*$ as they do for $\tilde n_i$. This is convenient as it is much less costly to compute: it does not require inverting the transform, which necessitates storing the full unitary transform at each flow time step. This also means that the accuracy of the final result is not limited by the resolution at which we store the unitary transform. A recent analytical study of operator spreading using flow methods can be found in Ref.~\cite{Hastings22}.

Although this operator is not a true local integral of motion, we can transform it into one simply by taking the long-time average:
\begin{align}
\overline{n}_i := \lim_{T\to \infty} \frac{1}{T}\int_0^T\mathrm dt\, e^{-iHt}n_i(l\to\infty)e^{iHt}.
\end{align}
To see that this is an integral of motion, notice that the off-diagonal terms of the form $\tilde c_i^\dagger c_j $ with $i\neq j$ -- along with higher order terms of the same form -- vanish under the time-averaging procedure, leaving behind only the terms which commute with the diagonal Hamiltonian. As a matter of fact, the operators $\tilde c_i^\dagger, \tilde c_i$ simply act as ladder operators for the eigenstates of the LIOMs (since both sets of operators are unitarily equivalent to the standard set of ladder operators), which count the number of particles in each ``quasilocal mode''. These eigenstates are also eigenstates of the Hamiltonian, which means that for an eigenstate of $H$ with energy $E$, $\langle E|c_i^\dagger c_j|E\rangle \propto \delta_{ij}$. Assuming the Hamiltonian has no degenerate states, these terms will give rise to oscillatory terms of the form $e^{i(E-E')t}$, which vanish when time-averaged. By contrast, the terms consisting of products of $\tilde n_i$ are unaffected by the averaging since they commute with the Hamiltonian. Together, this implies 
\begin{align}
\overline{n}_i =\sum_j \tilde{\alpha}^{(i)}_j \tilde{n}_j + \sum_{j,k} \tilde{\gamma}^{(i)}_{j,k} \tilde{n}_j \tilde{n}_k+ ...
\label{eq.fwd_liom_avg}
\end{align}
is a true integral of motion. Assuming that many-body localized systems satisfy a zero-velocity Lieb-Robinson bound \cite{Nachtergaele_2021,Sims_2013,Gebert_2016} -- that is, there exists constant $C$ such that $||[A(t),B]||_{\infty}\leq C$ for any $t$ -- it is easy to see that time averaging with an MBL Hamiltonian preserves quasilocality~\cite{Chandran+15}.

One of the aims of this work is to show that these operators are easier to compute and exhibit qualitatively similar behaviour to the LIOMs, making them ideal candidates for future studies of localisation in more computationally difficult scenarios, such as higher-dimensional systems or periodically driven models. We shall present numerical results to substantiate this claim in Sec.~\ref{sec.results}. 

\subsection{LIOM weight measure}

To investigate signs of a possible delocalisation transition using the LIOMs alone, we use the method developed in Refs.~\cite{Singh+21,Thomson+21} to estimate the transition point from the LIOMs. We define
\begin{align}
f_2 &:= \frac{1}{||n||}\left( \sum_j |\alpha^{(i)}_j|^2 + \sum_{j,k} |\beta^{(i)}_{j,k}|^2 \right),\\
f_4 &:= \frac{1}{||n||}
\left(\sum_{j,k} |\gamma^{(i)}_{j,k}|^2 + \sum_{j,k,p,q} |\zeta^{(i)}_{j,k,p,q}|^2 \right),
\end{align}
where $||n|| := \sum_j |\alpha^{(i)}_j|^2 + \sum_{j,k} |\beta^{(i)}_{j,k}|^2+\sum_{j,k} |\gamma^{(i)}_{j,k}|^2 + \sum_{j,k,p,q} |\zeta^{(i)}_{j,k,p,q}|^2$. In other words, the quantities $f_2$ and $f_4$ represent the fraction of the total weight of the operator contained by the quadratic and quartic components respectively, and give a measure for how important many-body effects are in the context of the LIOMs. In a strongly localised system with low entanglement, we would expect to see $f_2 \gg f_4$. On approach to the delocalised phase, the interacting terms become relevant and, \emph{heuristically}, we would expect to see 
\newtext{$f_4$ grow significantly and $f_2$ shrink significantly} as the highly entangled nature of the system requires the LIOMs to take on a highly non-trivial structure incorporating many-body interactions, this will result in observing $f_4\geq f_2$ on the thermal side of the transition.
It is important to note that these measures are somewhat heuristic in nature, and in particular that the limit of $f_4=1$ is unphysical. The TFE method breaks down in the delocalised phase, as discussed in Sec.~\ref{sec.TFE_diag}, and this results in the quartic components of the LIOM diverging. As the $f_n$ ratios are normalised, in this case we obtain $f_4=1$ and $f_2=0$. It must therefore be understood that the case of $f_4 \gg f_2$ represents a breakdown of the method which is linked to the proliferation of \emph{many-body} resonances that cannot be removed by the transform, indicative of a delocalised phase. We also wish to emphasise that in non-interacting systems, regardless of whether the system is localised or not we will always have $f_2=1$ and $f_4=0$, therefore care must be taken when interpreting this measure; it cannot simply be taken as a direct measure of how localised the system is, and should be understood as a heuristic only.
Note that we shall also define the ratios $\tilde{f}_2$ and $\tilde{f}_4$, which are the same measures but for the transformed number operator rather than the LIOM - as before, the tilde represents quantities computed in the diagonal basis.

\subsection{Truncated LIOM norm}

\begin{figure}[t]
    \centering
    \includegraphics[width=\linewidth]{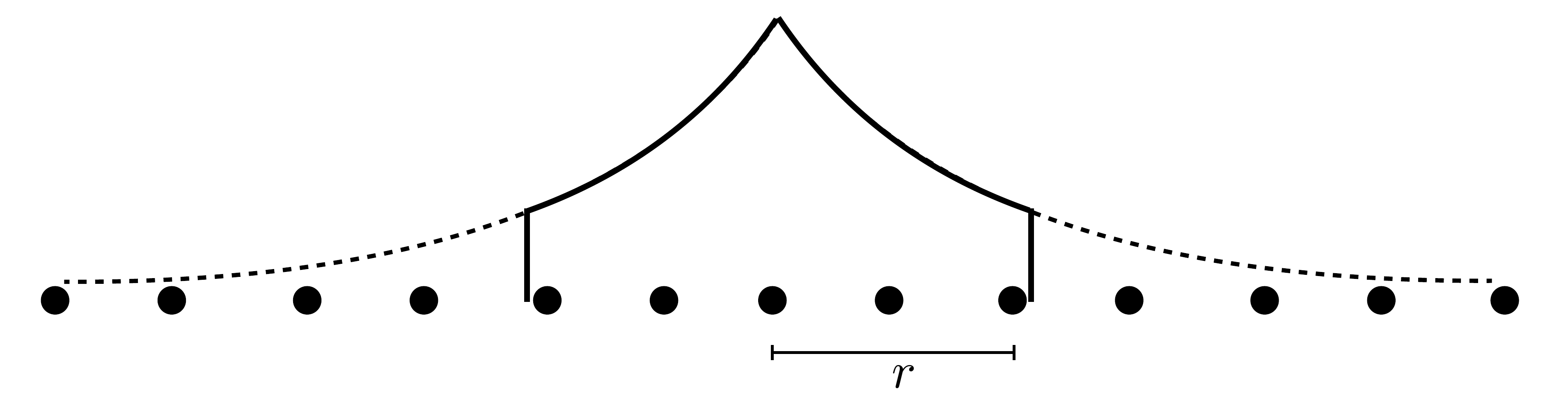}
    \caption{Pictorial representation of the truncated LIOM norm. Only terms that act inside of a certain radius $r$ are kept, and the tails that are cut off should be exponentially small.}
    \label{fig.liom_norm}
\end{figure}

We can also see the localisation in a measure that combines both the linear and the quadratic part of the LIOM. Let $O$ be an operator on the lattice with fermionic operator decomposition $O=\sum_{S} \alpha_S O_S$, where the sum is over all subsets $S$ of the lattice and $O_S$ are sums of products of creation and annihilation operators only acting on $S$. Define
\begin{equation}
    O|_{r,i}:=\sum_{S:d(S,i)\leq r}\alpha_S O_S,
\end{equation}
where $d(S,i)$ is the maximum distance between a site $i$ and the sites in $S$. We then define the locality of $O$ as
\begin{equation}
    L_O(r,i)=1-\frac{\left|\left|O|_{r,i} \right|\right|_F^2}{||O||_F^2},
    \label{eq.local}
\end{equation}
where $||\cdot||_F$ is the Frobenius norm. A visual representation is given in Fig.~\ref{fig.liom_norm}.

This locality measurement reduces to the one proposed in~Ref.~\cite{Goihl18} if the operators $O_S$ are orthogonal in the Hilbert-Schmidt inner product. Intuitively, if $O$ is well localised around the site $i$, restricting it to a certain radius around $i$ should not significantly change its norm, hence for large $r$ the quantity $L_O(r,i)$ should be small. In the following we always assume that $i=L/2$ is the central site of the chain and suppress this index for clarity. The Frobenius norm of a LIOM must be $2^{N-1}$, but since our method only gives us the linear and quadratic part of the LIOM, the norm of the operator constructed from these can deviate significantly from this value in the delocalised phase, as in this regime the LIOM might have significant weight on higher order terms. Computing this norm is computationally expensive, as it involves building the full LIOM as an exponentially large matrix rather than using the (polynomially large) tensor structure of Eq.~(\ref{eq.LIOM}), so instead of comparing the restricted LIOM to the full one, we instead compare it with the LIOM restricted to a maximum radius of $r=10$. Note that this definition involves $(2r+1)$ lattice sites, as we take $r$ sites either side of the LIOM centre, and as such even for $r=10$ is already computationally very costly to compute. By doing this, we measure how quickly in real space the LIOM converges to this particular restricted version instead of the full LIOM, but this has the same physical interpretation and (at least in the localised phase), terms beyond this radius should only have a very small weight. As we shall see, the results appear to give a consistent picture of LIOMs in disorder-free systems.

\section{Numerical results}
\label{sec.results}

\subsection{Linear potential}

\begin{figure}[t!]
    \centering
    \includegraphics[width=\linewidth]{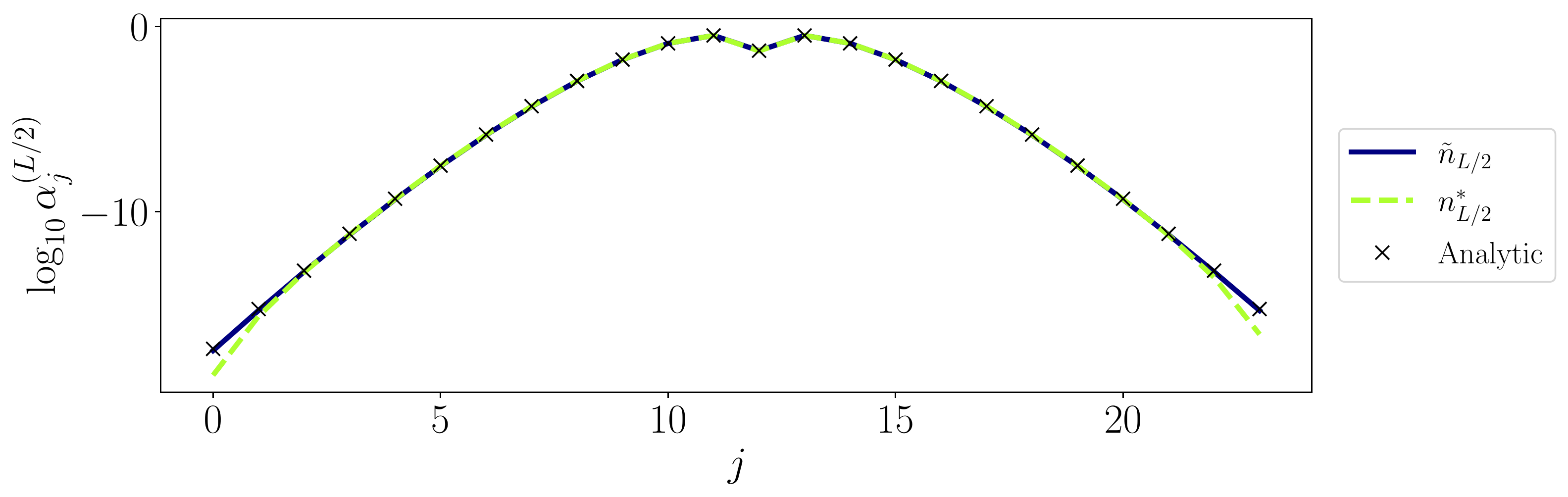}
    \caption{The diagonal, quadratic part of the $l$-bit ($\alpha^{(L/2)}_{j}$, blue) and transformed number operator ($\tilde{\alpha}^{(L/2)}_{j}$, green) located on the central site of a one-dimensional chain, as compared to the exact analytic solution (Eq.~(\ref{eq.bessel}), crosses). The $l$-bit agrees precisely with the analytical solution, while the transformed number operator exhibits small deviations at the edges of the chain. For clarity, we show the comparison for a single value of the field, $F=1.0$; other choices of $F$ show similar behaviour.}
    \label{fig.fwd_bck_liom}
\end{figure}

In this section, we present the numerical results obtained for the different choices of potential discussed in Section~\ref{sec.model}.
In the following, we set the on-site potential to be a purely linear slope, $h_i = F i$ where $F$ sets the strength of the linear slope and $i$ labels the lattice site. Following Ref.~\cite{vanNieuwenburg+19}, we would expect to see a transition as a function of $F/J$ for fixed interaction strength $\Delta_0 >0$. 

\subsubsection{Non-interacting 
systems}

We first verify the accuracy of our method by diagonalising the non-interacting Hamiltonian numerically using flow equations. As the system is non-interacting, this can be done exactly and we can compare with the known analytical solution for the form of the LIOMs. 
In Fig.~\ref{fig.fwd_bck_liom}, we show the diagonal terms $\alpha_j^{(i)}$ and $\tilde{\alpha}_{j}^{(i)}$ for the LIOMs and transformed operators respectively, in the case of a purely linear potential. We see that they are almost indistinguishable from each other, except from small deviations at the edge of the system, thus confirming that both operators evolve in very similar ways under the action of the quasi-local unitary transform used to diagonalise the Hamiltonian. (Note, however, that the off-diagonal terms of the transformed number operator are quantitatively quite different from those of the LIOM, as it does not commute with the Hamiltonian.)
Having confirmed that the method works as expected for the non-interacting system, we now move on to the many-body physics.

\subsubsection{Interacting system}

Now we switch interactions back on, and set 
$\Delta_0/J = 0.1$ such that any 3-body or higher-order terms in the diagonal 
Hamiltonian (Eq.~(\ref{eq.Hdiag})) can be considered negligible.
The first thing to note is that this method cannot fully 
diagonalise the many-body Hamiltonian due to the presence of a large number of many-body degeneracies. The transformed Hamiltonian takes the form 

\begin{equation}\label{eq:gamma_def}
\mathcal{H}(l\to \infty) = \tilde{\mathcal{H}} + \sum_{i,j,q} \Gamma_{i,j}^{q} \tilde{c}^{\dagger}_{i-q} \tilde{c}^{\dagger}_{j+q} \tilde{c}_i \tilde{c}_j.
\end{equation}
I.e., it contains a multi-particle hopping term which hops one particle `up' the linear slope by $q$ lattice sites, and another one `down' the linear slope by $q$ lattice sites, for a total change in the single-particle energies of $(-Fq +Fq) = 0$. This term couples degenerate states, and as such it cannot be removed by the flow equation method. This term allows for the movement of particles and strongly implies that localisation will not be stable to long times and large system sizes. It does not, however, rule out the existence of a seemingly localised phase that exhibits only very slow transport. 
We shall still refer to this as the `diagonal basis' as all single-particle off-diagonal terms have been removed, as have all multi-particle off-resonant terms, and as such the remnant off-diagonal terms carry very little weight.

\begin{figure}[t!]
    \centering
    \includegraphics[width=\linewidth]{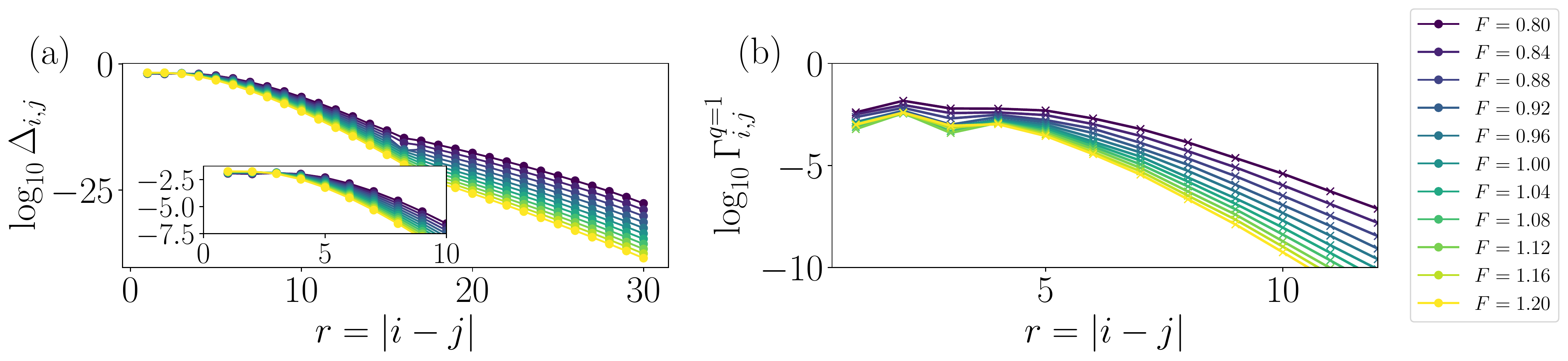}
    \caption{Results for a chain of length $L=36$ with interaction strength $\Delta_0=0.1$ for different values of the linear slope $F$. a) The LIOM interactions, $\Delta_{i,j}$ as a function of distance $r=|i-j|$. They display a faster than exponential decay at short distances (curved on a semi-log scale), with a crossover to what appears to be an exponential tail at large distances. The inset shows a zoomed in view at small distances $r$, highlighting the non-exponential decay. b) The coefficient of the remnant resonant terms that remain in the final Hamiltonian, $\Gamma_{i,j}^{q}$ with $|q|=1$, for the same parameters as the previous panel. This term is approximately an order of magnitude smaller than the maximum value of the $\Delta_{i,j}$ terms, but also displays a broad plateau that does not strongly vary with slope gradient.}
    \label{fig.linear_delta}
\end{figure}

We will now take a look at the behaviour of the LIOM interactions $\Delta_{i,j}$ and the resonant coefficient $\Gamma_{i,j}^{q}$ as defined in \cref{eq:Hdiag} and \cref{eq:gamma_def}, shown in Fig.~\ref{fig.linear_delta}. The interactions display an unusual structure that is more complex than the straightforward exponential decay seen in random systems. The LIOMs initially decay faster than exponential (i.e., curved on the semi-log scale of Fig.~\ref{fig.linear_delta}), before crossing over to behaviour consistent with a conventional exponential tail at large distances. By contrast, the LIOM interactions in a disordered system exhibiting conventional MBL decay almost perfectly exponentially with distance~\cite{Thomson+21}. 

This confirms that the many-body disorder-free scenario also behaves in a distinctly different way than conventional many-body localisation. Furthermore, we can see that the coefficient of the resonant terms $\Gamma_{i,j}^{q}$ (shown here averaged over $q=\pm1$; other values of $|q|>1$ are qualitatively similar, but quantitatively smaller) has a similar form to the LIOM interactions, with a broad plateau for short distances. Although this term is smaller than the LIOM interaction coefficient, its presence implies that resonant multi-particle processes leading to particle transport are present in this system, which likely leads to slow delocalisation on sufficiently long timescales. There is a weak variation as the slope gradient $F$ is increased, which acts to suppress these terms, however, the qualitative form is not strongly modified by a larger gradient. We note, however, that all of our analysis is conducted in the parameter regime where $F/J \lesssim 1$. In the regime of large gradients, $F/J \gg 1$, then it is reasonable to expect much faster decay of all parameters. In this parameter regime, the microscopic model is much closer to a classical Hamiltonian, as the linear slope is the largest energy scale in the problem and the hopping terms can be regarded as a small perturbation, implying that localisation is likely to be much more stable.

\begin{figure*}[t!]
    \centering
    \includegraphics[width=.85\linewidth]{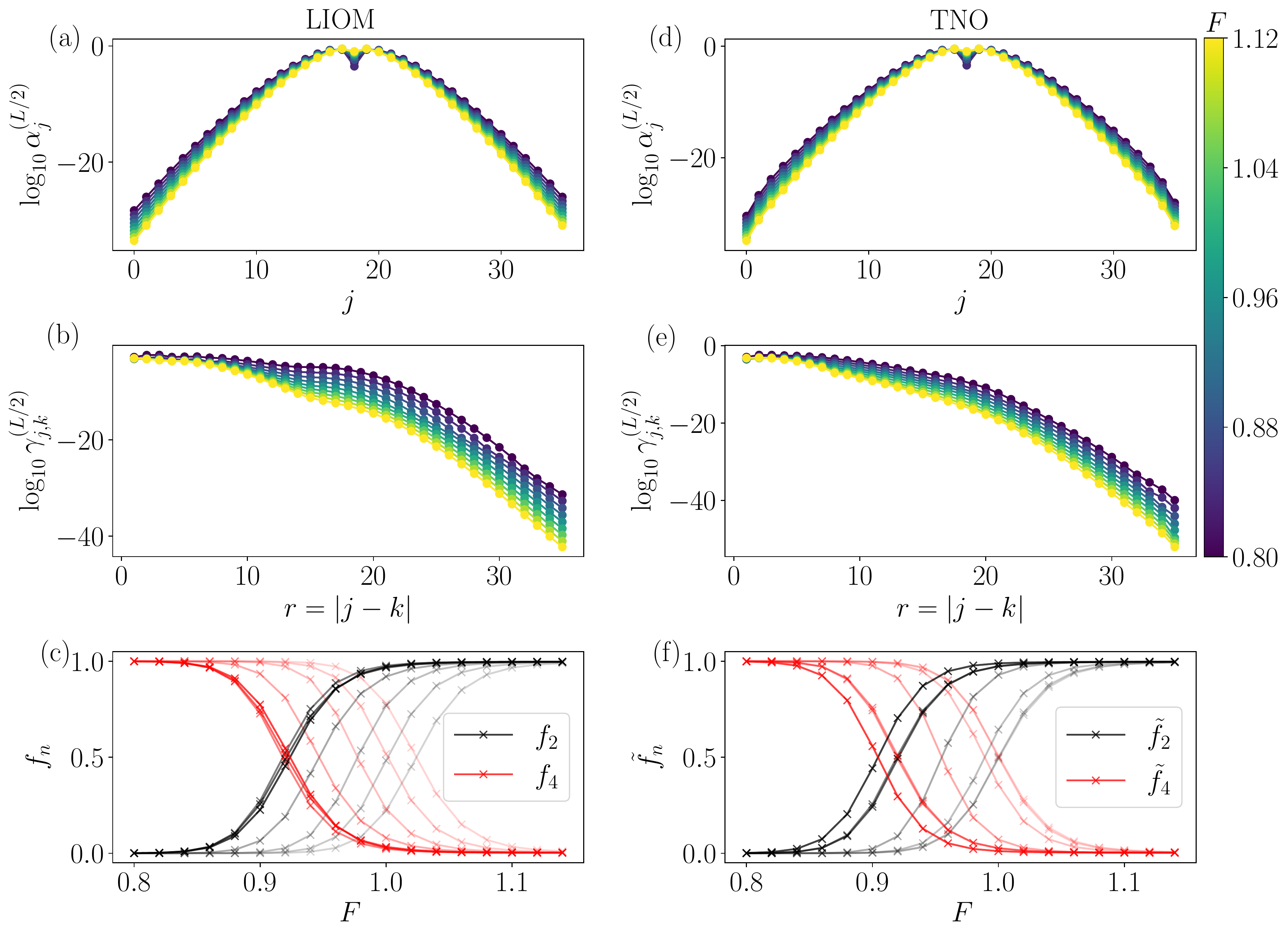}
    \caption{Results for a chain of length $L=36$ with interaction strength $\Delta_0=0.1$ for different values of the linear slope $F$. a) The quadratic part of the LIOM located on the central site of the chain. b) The diagonal quartic part of the LIOM. In contrast to disordered many-body systems, these LIOMs do not decay purely exponentially with distance. c) The ratios $f_2$ and $f_4$ for system sizes $L=8,10,12,16,24,30,36$ (light to dark), showing convergence with increasing system size. d) The quadratic part of the transformed number operator initially located at the centre of the chain. e) The diagonal quartic part of the transformed number operator (TNO). f) The ratios  $f_2$ and $f_4$ computed using the transformed number operator, for the same system sizes as in panel c).}
    \label{fig.linear}
\end{figure*}

We now turn to the  real-space support of the LIOMs themselves, as shown in Fig.~\ref{fig.linear}a-c). Panel a) shows the quadratic part of a LIOM located on the centre lattice site, for a variety of values of field strength $F$. In the present scheme, we do not implement normal-ordering corrections in the flow equations (see Refs.~\cite{Kehrein07,Wegner06} for further information), and as such the $\alpha^{(i)}_j$ coefficients are identical to those of the non-interacting system. Panel b) shows the diagonal part of this two-body contribution to the LIOM, which exhibits a clearly non-exponential decay with distance. For small values of $F$, the interacting part of the LIOM shown in Fig.~\ref{fig.linear}b) displays a lengthy plateau that decays only very slowly with distance, before crossing over to a faster exponential-like decay at large distances. This is interesting as it shows that although the quadratic component of the LIOM is always localised (as in the single-particle model), the interaction terms can become highly extended. This suggests that the interaction terms could be responsible for driving a delocalisation transition, or at the very least that one might expect spatially separated LIOMs to have a larger overlap than in conventional disorder-driven MBL. This is consistent with the faster-than-logarithmic growth of the entanglement entropy that has already been seen numerically~\cite{Schulz+19}, which also implies an overlap between LIOMs that is not exponentially small, as in conventional disorder-driven MBL.

At the smallest values of $F$ note that there is a visible divergence of $\gamma_{j,k}^{(L/2)}$ at small distances, due to the convergence of the flow becoming slow (as $F/J \ll 1$) and the large number of degeneracies in the many-body spectrum leading to the quartic terms in the LIOM diverging. This suggests that the system cannot be diagonalised in this manner, and may become delocalised at the smallest values of $F/J$ shown here. 
We can investigate this more quantitatively through the ratios $f_2$ and $f_4$, defined previously.
Due to the divergence visible in Fig.~\ref{fig.linear}b) for small values of $F$, the term $f_4$, which takes into account the relative weight of the quartic terms in the LIOM, will tend towards one, and $f_2$ will vanish. In principle, the crossover from $f_2=1, f_4=0$ to $f_2=0,f_4=1$ can be used to estimate the transition point, although it should be noted that what this measure really probes is the point at which degeneracies in the Hamiltonian lead to diverging terms in the LIOM and the method breaks down. This is likely to be close to the phase transition but care should be taken when interpreting these results. The ratios $f_n$ are shown in Fig.~\ref{fig.linear}c) for various different system sizes. We see that there is a significant finite-size effect, with the crossover to localised behaviour shifting to smaller and smaller values of the slope $F$ as the system size is increased. While it is tempting to conclude that this implies localisation is more stable for larger system sizes, this interpretation is not supported by the full data. Rather, it is more likely that this effect simply indicates that these systems are subject to severe finite size effects; for example, in Fig.~\ref{fig.linear}b), 
we can see that systems of size $L \leq 24$ are within the plateau region of the LIOMs, and it is only on larger length scales that the quasi-exponential tails become visible. For systems large enough to resolve this tail, the $f_n$ ratios appear to converge.

In Fig.~\ref{fig.linear}d-f), we show the same quantities computed using a transformed number operator on the central lattice site, rather than the true LIOM. All features are qualitatively identical, with only small quantitative changes, although in general it seems that the transformed number operator is marginally more local than the true LIOM, with all quantities exhibiting a slightly faster decay. We remind the reader that this observable is far less computationally challenging to compute, and the strong resemblance to the behaviour of the true LIOM is an encouraging sign that it captures the same essential physics at a much lower computational cost. This quantity could be a promising candidate for future study in more demanding regimes, such as two-dimensional systems or periodically driven models~\cite{Thomson+20c}.

\begin{figure}[t!]
    \centering
    \includegraphics[width=1.1\linewidth]{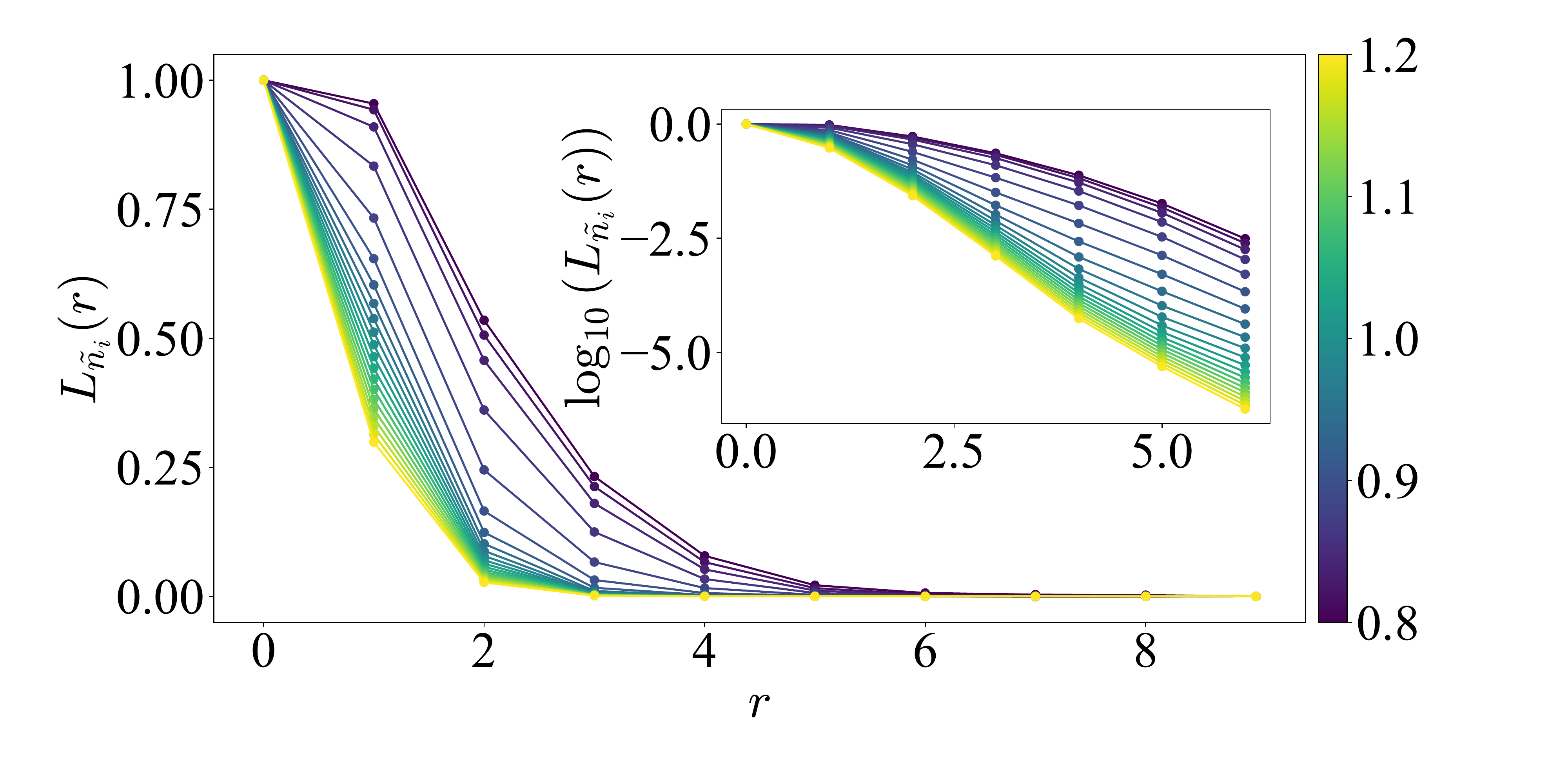}
    \caption{Locality of the central LIOM in a system of size $L=24$, the different lines are different values of the slope, from $F=0.8$ to $F=1.2$ at intervals of $0.02$. The insert shows the same quantity on a logarithmic scale, where purely exponential decay would appear as a straight line. It is clear that the truncated norm of the LIOMs of the disorder-free system do not decay exponentially. }
    \label{fig.locality_linear}
\end{figure}
We can also compute the locality of the central LIOMs via their truncated norm. %
In Fig.~\ref{fig.locality_linear}), one can see this quantity as a function of the distance from the center for values of the slope between $F=0.8$ and $F=1.2$ at intervals of $0.02$. For small slopes,the locality decays slowly at small distances before starting to drop towards zero, while for large slopes the locality immediately drops quasi-exponentially, signalling that the LIOM is strongly localised. This echoes the behavior of the decay of the quartic part of the LIOM observed in 
Fig.~\ref{fig.linear_delta}: while the rapid decay of the quadratic part of the LIOM attenuates the effect, the plateau observed in the quartic part is still visible in the overall decay of the LIOM norm. These results again demonstrate that the real-space support of the LIOMs does not behave as expected in conventional MBL, as it decays in a non-exponential matter. This slow decay of the LIOM strongly implies that nearby LIOMs will have an enhanced overlap as compared with the case of conventional disorder-driven MBL, which could account for the fast growth of entanglement entropy seen in earlier works~\cite{Schulz+19}.

\subsection{Quadratic potential}

We can now add some curvature to the potential in order to reduce degeneracies of the many-body spectrum, as in Ref.~\cite{Schulz+19}. We now set the on-site potential to be $h_i = F i + \alpha (i/L)^2$, and repeat the analysis from the previous section. 

\subsubsection{Non-interacting systems}

We again start by plotting the real-space support of the LIOMs in the non-interacting system. With the addition of the curved potential, there is no analytic solution, however,
the results of 
Ref.~\cite{Schulz+19} would suggest that the system is more strongly localised in the presence of the curvature, and therefore we might expect to see the LIOMs decay more quickly in real space. 

\begin{figure}[t]
    \centering
    \includegraphics[width=\linewidth]{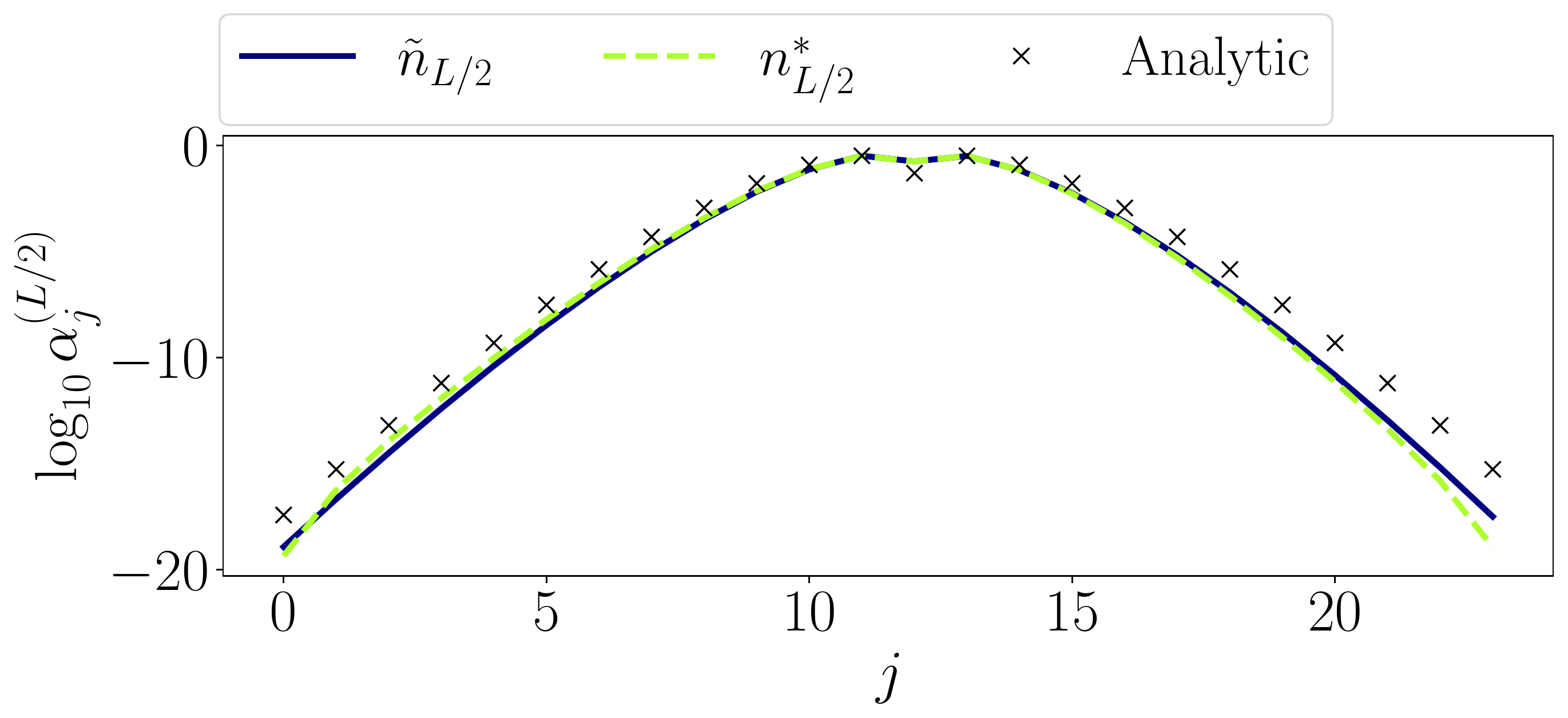}
    \caption{A comparison of a LIOM computed for the quadratic potential with $\alpha = 5.0$ (purple line) with the analytical solution of the purely linear potential (crosses), with system size $L=36$. The LIOM is more strongly localised in the presence of the curvature. The green line shows the diagonal part of the transformed number operator, as in Fig.~\ref{fig.bessel}, which again agrees very closely with the behaviour of the true LIOM except for small deviations at the edges of the chain.}
    \label{fig.curved_nonint}
\end{figure}

The results are shown in Fig.~\ref{fig.curved_nonint}, and indeed this is precisely what we see. The black crosses represent the analytical solution in the $\alpha=0$ limit (as shown in Fig.~\ref{fig.bessel}), while the lines represent the numerical data for the curved potential with $\alpha = 5.0$). We see that the LIOMs decay much faster in the presence of this curvature than they did in the purely linear potential. Curiously, the oscillatory behaviour close to the centre of the LIOM is also suppressed. This confirms that indeed the addition of a quadratic potential enhances localisation of the single-particle wavefunctions, but what effect does it have on the interaction terms of the LIOM? We now turn to the many-body case in order to find out.

\subsubsection{Interacting systems}

\begin{figure}
    \centering
    \includegraphics[width=\linewidth]{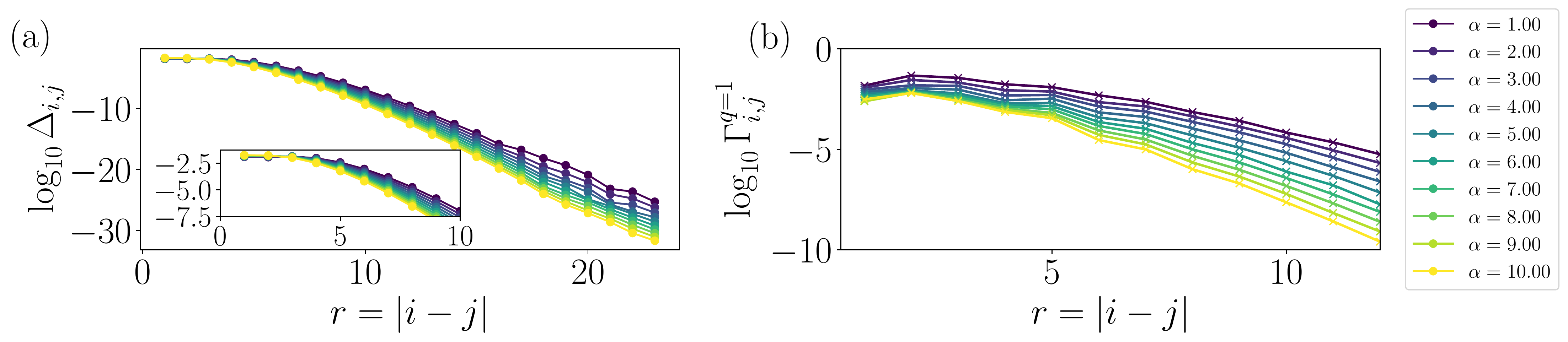}
    \caption{a) LIOM interactions extracted from the diagonal Hamiltonian (Eq.~(\ref{eq.Hdiag})) in the case of a linear potential with weak curvature, controlled by the parameter $\alpha$. Results are shown for $L=24$ at a fixed slope $F=0.8$. Qualitatively similar results are observed for other values of the slope. We see that increasing $\alpha$ has a qualitatively similar effect to increasing $F$ in the purely linear case, i.e., it leads to a faster decay of the LIOM interactions. b) The coefficient of the remaining resonant terms in the final transformed Hamiltonian. Here we can see that the curvature plays a significant role in removing these resonances as the curvature parameter $\alpha$ is increased.}
    \label{fig.curved_delta}
\end{figure}

We can now add the interactions back in, again using $\Delta_0=0.1$, and again we will look at the real-space decay of the LIOM interactions $\Delta_{i,j}$, the real-space decay of the LIOMs themselves, and the ratios $f_2$ and $f_4$ used to indicate (de)localisation. By adding curvature which breaks the degeneracies of the many-body states, we find that the resulting Hamiltonian is much closer to being truly diagonal, with fewer resonant terms surviving until the end of the flow.
Taking the interactions first [Fig.~\ref{fig.curved_delta}a)], we see a smoother decay than in the case of the purely linear potential. By comparison with the purely linear potential, it is clear that the curved potential is significantly more localised for all values of $F/J$, as 
one might reasonably expect. In Fig.~\ref{fig.curved_delta}b), 
we see that the curvature has a significant effect on the presence of off-diagonal resonant terms that remain in the final Hamiltonian, being much more effective at suppressing them than simply increasing the slope $F/J$. This suggests that for large curvature, any remaining resonant hopping terms will be so small that their effects are unlikely to be seen except on extremely large timescales.

\begin{figure*}
    \centering
    \includegraphics[width=0.85\linewidth]{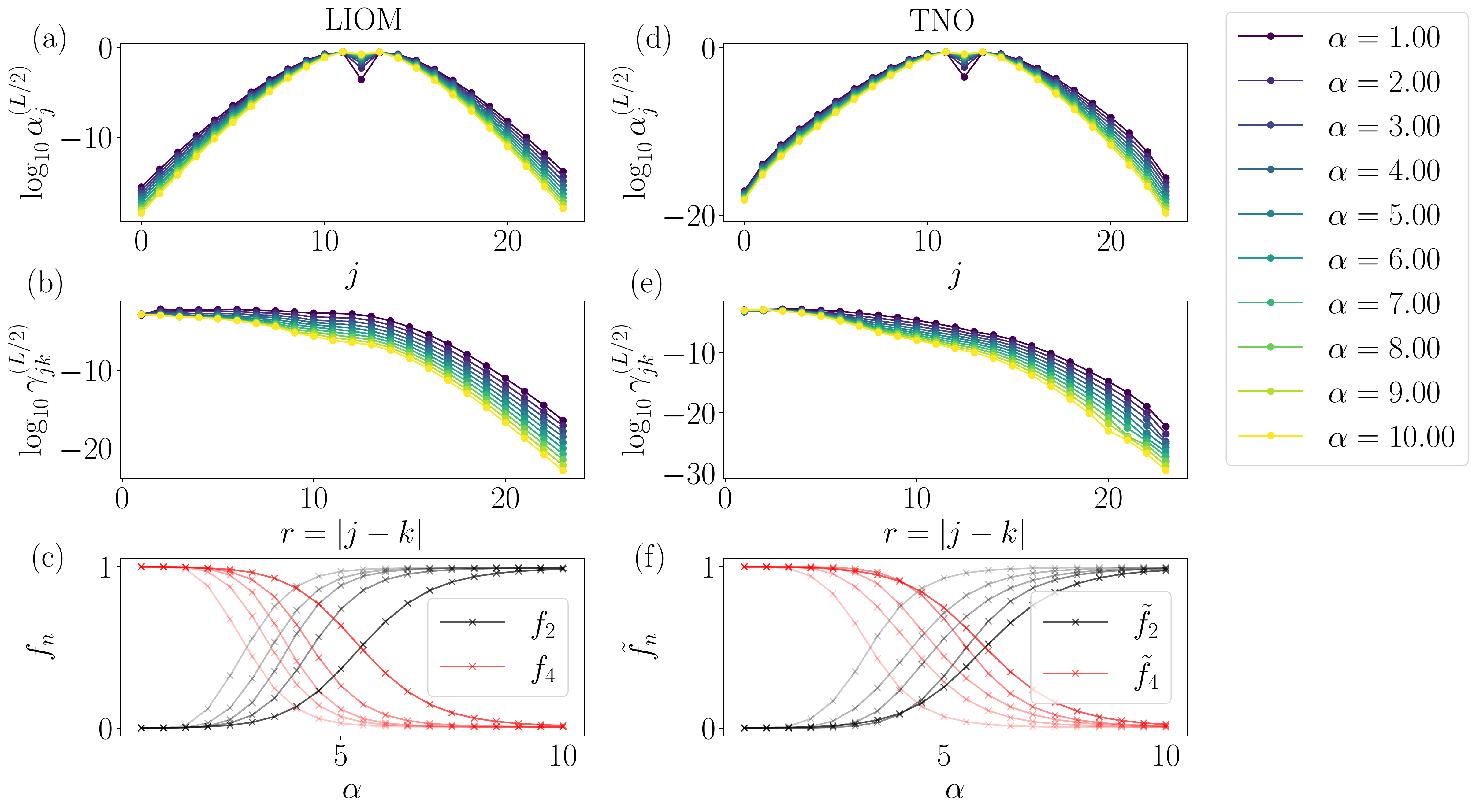}
    \caption{The same quantities as shown in Fig.~\ref{fig.linear}, but now computed for the potential with weak curvature. Unless stated, the results are for system size $L=24$ with fixed slope $F=0.8$. a) The quadratic part of the LIOM located on the central site of the chain. b) The diagonal quartic part of the LIOM. c) The ratios $f_2$ and $f_4$ for system sizes $L=8,10,12,16,24$ (light to dark). d) The quadratic part of the transformed number operator initially located at the centre of the chain. e) The diagonal quartic part of the transformed number operator. f) The ratios  $\tilde{f}_2$ and $\tilde{f}_4$ computed using the transformed number operator, for the same system sizes as in panel c).}
    \label{fig.curved_bck}
\end{figure*}

We now turn to the LIOMs themselves. The results are shown in Fig.~\ref{fig.curved_bck}a-c), where we compute the same quantities as in Fig.~\ref{fig.linear}, but now for a fixed slope $F=0.8$ and for varying values of the curvature parameter $\alpha$. For small values of the curvature the system behaves largely the same as in the purely linear case, with single-particle LIOMs which exhibit oscillatory behaviour near the central site and a many-body contribution which has a lengthy plateau before a slow crossover to exponential decay at large distances. As the curvature is increased, however, the Bessel-function-like behaviour of the LIOM is suppressed and the single-particle component decays much faster. The many-body contribution still displays a plateau, but it is significantly weakened in magnitude and exhibits a slow decay. This can be more precisely quantified using the $f_n$ ratios, shown in Fig.~\ref{fig.linear}c), where we see again that there is a strong finite-size dependence and a broad crossover from localised to delocalised behaviour at small values of $\alpha$. The effect of increasing the system size is rather different in the case, with larger values of $L$ tending towards delocalisation. This may seem surprising, but it is a consequence of the $1/L^2$ term in the potential which causes the curvature to become weaker with increasing system size, eventually vanishing in the $L \to \infty$ limit, where we would expect to recover the result from the purely linear slope.

In Fig.~\ref{fig.curved_bck}d-f), we show the same quantities computed for the transformed number operator, where we note that the qualitative behaviour is again the same as that of the true LIOMs. Interestingly, this operator seems to pick up the effect of the curvature more strongly than the LIOM, with a clear asymmetry visible in Fig.~\ref{fig.curved_bck}d) that much less visible in the corresponding Fig.~\ref{fig.curved_bck}a), although a careful examination reveals that the asymmetry is indeed present in both. The decay visible in Fig.~\ref{fig.curved_bck}d) is faster than that in Fig.~\ref{fig.curved_bck}b), again consistent with the findings in the previous section that the transformed number operator decays more quickly than the true LIOM.

\begin{figure}
    \centering
    \includegraphics[width=1.1\linewidth]{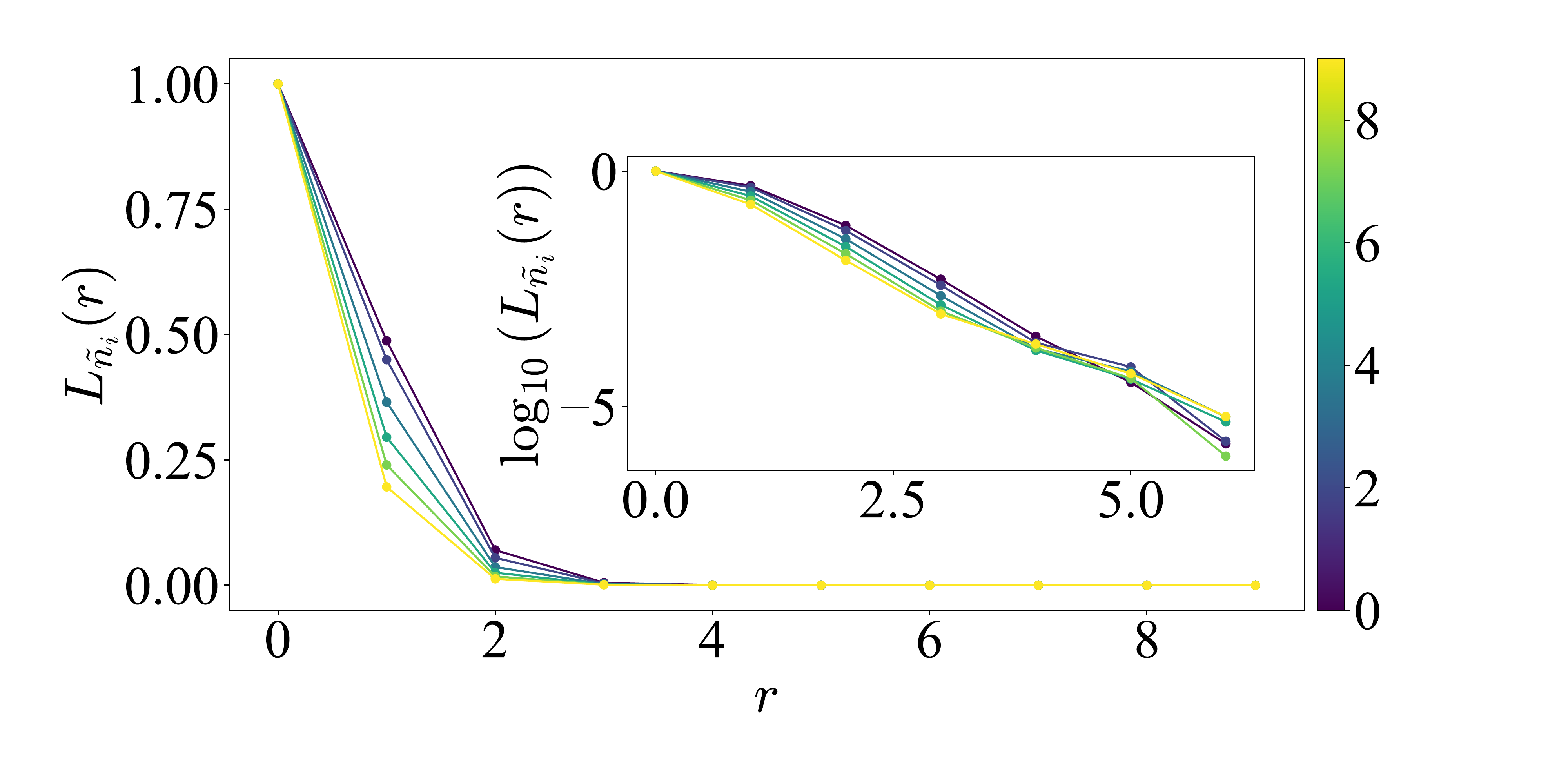}
    \caption{Truncated LIOM norm for fixed slope $F=1.0$ and varying curvature $\alpha$. System size: $L=24$.}
    \label{fig.locality_curved}
\end{figure}

Finally, we can look at the truncated LIOM norm for the curved potential. Figure \ref{fig.locality_curved} shows the LIOM norm for a fixed slope of $F=1.0$ and increasing curvature. It is clear that the addition of the curvature makes the LIOM more localised, but once again even at substantial curvature a short distance non exponential behavior is preserved due to the failure of the quartic part to decay exponentially.

\begin{figure}
    \centering
    \includegraphics[width=\linewidth]{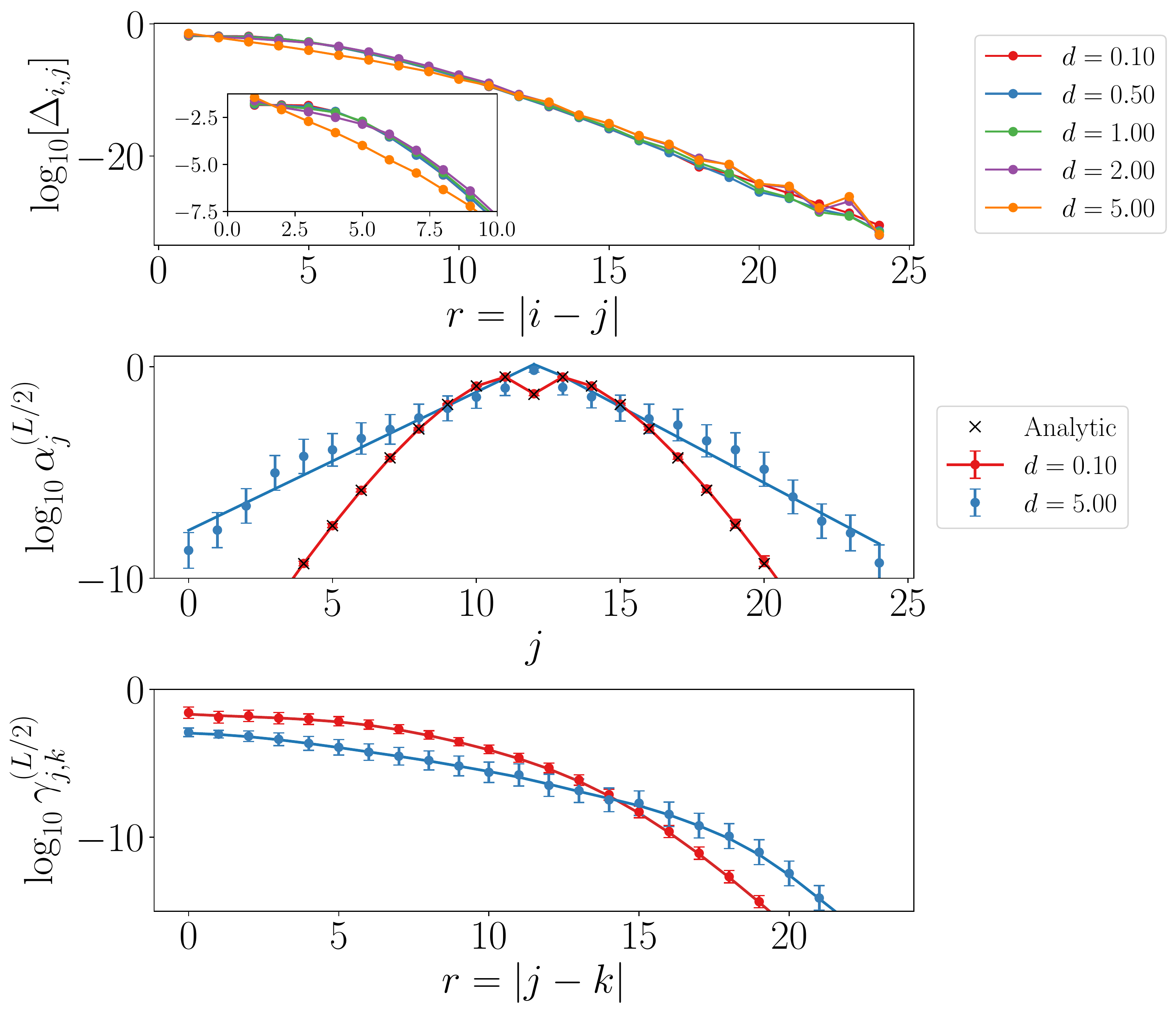}
    \caption{The LIOMs obtained for a linear potential with slope $F=1.0$ with added disorder of strength $d$, shown for a system size of $L=25$. a) The decay of the LIOM interactions, showing the typical (median) decay (with $N_s=256$ disorder realisations), highlighting the clear visible difference in the short-range behaviour between weak disorder and strong disorder. In the latter case, the decay at short distances is clearly exponential, whereas in the former case it exhibits a short plateau. Error bars are omitted for clarity. b) The diagonal quadratic component of the LIOMs, averaged over $N_s=512$ disorder realisations, with the analytical result of Eq.~(\ref{eq.bessel}) indicated by black crosses. Error bars indicate the mean absolute deviation. Weak disorder does not strongly modify the properties of the LIOMs, however, strong disorder leads to a qualitative change in behaviour from a Bessel function to an approximately exponential decay. (Here we only show two values of disorder for clarity.) c) The diagonal quartic component of the LIOMs ($N_s = 512$) displays similar behaviour, with weak disorder behaving qualitatively the same as the clean system, while strong disorder leads to a smaller tail that decays more quickly, consistent with exponential decay.}
    \label{fig.disordered}
\end{figure}

\subsection{Weakly disordered potential}

Finally, we examine the effect of adding a weak additional random potential on top of the linear potential. The results are shown in Fig.~\ref{fig.disordered} for two disorder strengths, $d=0.1$ and $d=5.0$, averaged over $N_s \in 
\{256,512\}$ 
disorder realisations.
Taking the LIOM interactions first [Fig.~\ref{fig.disordered}a)], 
we can immediately see a dramatic difference in the decay of the interactions as disorder is increased. At weak disorder, we see the by now familiar non-exponential decay that characterises the linear slope, with a short plateau at small distances. For strong disorder, however, the LIOM interactions decay at short distances almost perfectly exponentially (a straight line on a semi-log plot, most clearly visible in the inset). The long distance tails are very similar in both cases, as the influence of the linear potential is still felt at long distances, but the short-range behaviour in the presence of strong disorder is much more in line with previous studies of disorder-driven localisation. We do not show the remaining resonant terms $\Gamma_{i,j}^q$ for the disordered model, as for any $d > 1$ they are essentially zero to within machine precision, as the disorder suppresses all degeneracies and so the flow equation transform is able to remove all such off-diagonal terms without difficulty. This again highlights the very different way in which the disordered and disorder-free potentials behave, and from this we conclude that disorder is more effective at stabilising localisation than either of the disorder-free potentials studied in this work.

Turning to the LIOM itself, we can also clearly see that weak randomness does essentially nothing to the form of it, and the quadratic component agrees very closely with the analytical form of Eq.~(\ref{eq.bessel}) [black crosses in Fig.~\ref{fig.disordered}b)]. This makes sense, as such weak disorder is only a minor perturbation on top of the linear slope, and is not strong enough to add any new resonances into the system or otherwise change its behaviour in any qualitative manner. Strong disorder, on the other hand, dramatically modifies the LIOM such that both the quadratic and quartic components exhibit behaviour consistent with exponential decay at short distances, as seen in conventional models of disorder-driven MBL. In the presence of such strong disorder, the tail of the interacting component of the LIOMs is smaller and decay faster than in the purely linear or weakly disordered potentials. 
It is worth emphasising that even for the strongest disorder which we consider here ($d=5.0$), the linear slope still plays a significant role at large distances, and as such we do not recover precisely the properties of a random system in the absence of a linear potential. Hints of this can be seen in Fig.~\ref{fig.disordered}c) where the exponential decay crosses over to another functional form at large distances.

\section{Entanglement Entropy}

So far we have computed the properties of local integrals of motion in a variety of systems subject to potentials which are close to the Wannier-Stark ladder, and we have suggested that the non-exponential decay of the LIOMs is responsible for the faster-than-logarithmic growth of entanglement entropy seen in Ref.~\cite{Schulz+19}. Here we make this statement a little more precise.

The logarithmic growth of entanglement with time seen in many-body localised systems is commonly argued to arise from the exponential decay of the overlap of distant LIOMs~\cite{Bardarson+12,SerbynPapicAbaninPRL13}. 
This, however, does not take into account the effect of a plateau, which we expect to strongly enhance the overlap between LIOMs separated by short distances, potentially allowing transport throughout the system.
It has previously been suggested that the \emph{entanglement entropy} in a purely linear potential grows \emph{faster} than logarithmically~\cite{Schulz+19}, implying a larger-than-exponential overlap between the LIOMs. This is consistent with our findings that the LIOMs do not decay in an exponential manner. 

In fact, for MBL featuring disorder
it has been shown that the disorder-averaged 
entanglement entropy 
can grow at most logarithmically in 
time~\cite{Kim14}, starting from product states. These upper bounds are proven by
decomposing the unitary reflecting time evolution
generated by the Hamiltonian 
as an operator acting only on two parts, each with
respect to which notions of entanglement are
considered, and another unitary that is supported on the
vicinity of the boundary between those parts~\cite{PhysRevLett.97.150404}. Following the 
logic of the argument of Ref.~\cite{Kim14},
decomposing the Hamiltonian into LIOMs (the
strength of which are sufficiently small), 
it is plausible that for LIOMs
featuring a non-exponential decay as featured here
one finds an upper bound for the growth of the
entanglement entropy that scales faster than 
logarithmically in time. 
the findings of Ref.~\cite{Schulz+19}.

To see this, we note that it has been shown in Ref.~\cite{Kim14} that the entanglement growth in a many-body localised system following a quench from a product state can be upper bounded by a function of the form
\begin{align}
 S(t) \leq C_1 r + C_2 \exp(-r/\xi) t,
 \label{eq.entropy}
\end{align}
where $r$ is some length scale that can be freely chosen. By choosing $r = \xi \log(t)$, one can recover the familiar logarithmic growth of the entanglement entropy with time. The key point is that this choice of $r$ completely cancels the linear growth in the second term of Eq.~\ref{eq.entropy}: the exponential dependence on $r$ here comes directly from the exponential decay of the LIOMs. If we instead assume a more relaxed form of the LIOMs, we arrive at the 
\newtext{expression}
\begin{align}
    S(t) \leq C_1 r + C_2 f(r) t,
\end{align}
where $f(r)$ encodes the form of the LIOM. Based on our numerical results which show a plateau, we can gauge the effect of this by crudely estimating $f(r)$ to be piecewise constant and composed of a constant plateau at short distances, followed by an exponential decay. We stress that while this picture is undoubtedly an oversimplification, it serves to illustrate the effect of LIOMs which do not decay exponentially as
\begin{align}
    f(r) = 
    \begin{cases}
        const., &  \text{if}\ r < R ,\\
        e^{-r/\xi}, &  \text{otherwise},
    \end{cases}
\end{align}
which in turn leads to
\begin{align}\label{eq:entropy}
    S(t) \leq
    \begin{cases}
        \mathcal O(t), &  \text{if}\ t < \exp(R/\xi) ,\\
        \mathcal O(\log(t)), &  \text{otherwise}.
    \end{cases}
\end{align}
%
This analysis suggests that a slower-than-exponential decay of the LIOMs can lead to a linear increase of the entanglement entropy at short times, with a crossover to logarithmic growth at late times. This may in fact be visible in the data of Ref.~\cite{Schulz+19}. As the `plateau' region of the LIOM is suppressed, the crossover from linear to logarithmic growth is moved to shorter and shorter times. By increasing the gradient of the linear field, or by lifting the degeneracies through the addition of weak curvature or weak disorder, this plateau may be weakened and conventional MBL behaviour eventually recovered.


We conclude this section by formulating two questions which are crucial to understand whether clean Hamiltonians can exhibit true MBL or whether some explicit (i.e., a random potential) or implicit (i.e., as seen in lattice gauge theories) form of disorder is necessary. 
First, we must ask how $R$ behaves as a function of the system size $L$. In particular, even a very slow growth of the order $R(L)\sim\log(L)$ implies the entanglement entropy will grow linearly for a time algebraic in the system size, e.g., $t\sim L^{1/\xi}$. 
This connects to the second question: under which circumstances is one willing to call such systems localized? 
To see the implications of these questions let us first assume $\xi\lesssim 1$ and $R(L)\sim\log(L)$. Any such system should not be considered localized as the entropy will grow extensively. To see this we recall that in the thermodynamic limit it is crucial to scale $L$ and $t$ in the right order, namely $\lim_t\lim_L$. 
For $\xi>1$ the question becomes more subtle as in this case, following Eq.~\eqref{eq:entropy}, the entanglement will exhibit both a linear growth regime (at short times) and a logarithmic growth regime (at long times). 
Even for systems displaying MBL, it is natural that the entanglement entropy has a linear onset before crossing over into a logarithmic regime (except in the limit of infinitely strong disorder). It is thus interesting to understand the properties of the short-time linear growth regime in more detail in order to establish whether a system can be considered many-body localised. If the timescale at which the crossover to logarithmic growth occurs at a time scaling as a constant in the system size $t \propto L$, this may be regarded as MBL perturbed away from the idealised infinite-disorder limit. 
The same reasoning may be extended to the case when the crossover happens at a time $t\propto\log(L)$, which still corresponds to a state that after any polynomial time (in $L$) has only a logarithmic (in $L$) entanglement entropy. The situation is different in case of a crossover time $t \propto L^{1/\xi}$. In this setting, the underlying state would have an algebraic (in $L$) entanglement entropy  after a linear time, which would break the intuition that many body localised systems have low entanglement.

\section{Discussion and conclusions}
\label{sec.conclusion}

We have computed the local integrals of motion for interacting fermions subject to both disorder-free and weakly disordered potentials. In the case of disorder-free potentials, we have shown that the LIOMs and their localisation properties are qualitatively different from those of conventional many-body localisation seen in random or quasi-random potentials. Our results demonstrate that the LIOMs can be broken into single-particle and many-body components, and despite the single-particle terms decaying faster than exponentially, at weak values of the slope $F/J$ the many-body components can extend significantly throughout the system, driving a delocalisation transition. This does not necessarily imply ergodicity, as exotic non-ergodic extended states have been shown to exist in close proximity to localised phases~\cite{Kravtsov+15}. This is a topic that should be further explored in future work.

It should be noted, however, that our localisation measures are based upon the real-space support of individual LIOMs, but in a real many-body system the \emph{overlap} between spatially separated LIOMs also plays a role in how quickly the system will delocalise when prepared in some initial arbitrary excited state. 
In particular, for systems with no -- or only weak -- disorder, we find that the interacting component of the LIOMs exhibits a plateau around their center site before decaying in a faster than exponential manner. 
In contrast, at sufficiently strong disorder the LIOMs do not show any such plateau and the exponential decay sets in immediately (c.f.\ Fig.~\ref{fig.disordered}).
We believe that this distinction is a key feature for the difference in the dynamics of these systems.

One particularly striking feature of our results is that increasing the strength of the curvature gradually from zero has a very similar effect to simply increasing the gradient of the linear potential. Importantly, the qualitative form of the LIOMs and in particular the existence of a plateau remains valid. Treating this as a witness for delocalisation, we believe that the quadratic potential ultimately inherits the instability of the linear potential, although the lifting of some of the degeneracies weakens this effect, meaning it may only show up in the long time limit. This is in contrast to what may have been expected based on previous studies where observables were found to behave in a manner consistent with conventional disorder-driven MBL~\cite{Schulz+19}. We interpret our result as strong evidence that in fact localisation is not significantly more stable in the curved potential than in the linear potential with a large gradient, and that for sufficiently large systems and long times, the system may ultimately thermalise.

This presents a clear challenge to the prevailing wisdom of disorder-free many-body localisation. If weak curvature stabilises localisation, as suggested in Ref.~\cite{Schulz+19}, then we would expect to see the LIOMs approach the conventional exponentially-localised form as the curvature is increased from zero. However, we find that both the quartic component of the LIOMs and their interactions retain clear signs of a plateau at short distances, which is only weakly affected by increasing the curvature further. While it is true that the long-distance tails of the LIOMs decrease in a manner consistent with an exponential decay, the non-exponential behaviour at short distances is unchanged by the addition of weak curvature. In contrast, the quadratic component of the LIOMs changes dramatically when true random disorder is added, approaching the conventional exponentially decaying form seen in systems with disorder alone.
These observations can be heuristically explained by considering the amount of degenerate states that are eliminated by choosing a given potential, as discussed in Section~\ref{sec.degen} and which we briefly recap here. The linear slope ensures that states that are separated by a single hop are never degenerate: They are separated by the energy cost $F$. This means that the off-diagonal couplings between these states can be treated perturbatively. The same is not true for states separated by multiple hops, which play a significant role at higher orders in perturbation theory. The highly degenerate nature of the spectrum is responsible for the poor localisation properties of the linear slope potential. Adding some weak curvature to the linear potential improves the situation by removing many of the higher order degeneracies, but nevertheless a similar argument applies and the many-body eigenvalues of the system in the presence of the curved potential can still take only quadratically many values which are to be shared by exponentially many states, as argued in section \ref{sec.degen}. 

\newtext{Our work complements the recent results of 
Ref.\ \cite{Halimeh+22c} in which a large class of symmetric Hamiltonians which includes the pure linear slope examined here is shown to exhibit transport. Here, we show numerically that breaking the symmetry with a harmonic potential does not qualitatively change the physics, and conjecture that both of these cases might be explained by the proliferation of resonances as described above.}
Based on the results presented in this manuscript, it would also be interesting to apply similar locality measures to the LIOMs in disorder-free systems in two dimensions, a regime which has seen recent numerical evidence for MBL~\cite{Wahl2017,Thomson+18,Augustine2DMBL,Chertkov2DMBL,Decker2DMBL} despite its instability argument~\cite{RoeckImbrie,deroeck2017}. Another interesting setting for this technique would be to study disorder-free localisation in lattice gauge theories~\cite{Smith+17,Brenes+18,Karpov+21,Halimeh+22a,Halimeh+22b,Lang+22,Chakraborty+22} where the underlying localisation method is quite different and there is no direct link to the concept of local integrals of motion in the way that they are defined here, although it seems reasonable to assume that the conserved charges of the theory must play a similar role and may be able to be expressed in terms of a quasi-local unitary transform applied to some non-trivial microscopic operator. This is particularly interesting because in such systems there is an emergent form of disorder, which may lead to more stable localisation than in the Wannier-Stark case studied in this manuscript. At the end of the day, it may be \emph{quantum simulations}~\cite{CiracZollerSimulation} 
that will ultimately be able to decide on the
stability of these various readings 
of localization discussed here. This strategy
of settling questions in quantum many-body
theory based on the outcome of quantum simulations constitutes an intriguing perspective.

\section*{Acknowledgements}
On the day this work was originally posted,
a related work 
(Ref.~\cite{Halimeh+22c}) was published that explores the absence of localization in interacting spin chains with a discrete symmetry.
This project has received funding from the European Union’s Horizon 2020 research and innovation programme under the Marie Skłodowska-Curie grant agreement No.~101031489 (Ergodicity Breaking in Quantum Matter), grant agreement No.~817482 (PASQuanS), and the Deutsche Forschungsgemeinschaft 
(CRC 183 and FOR 2724). We also acknowdge funding
from the BMBF (FermiQP and MUNIQC-ATOMS) \newtext{and the ERC (DebuQC).}
SJT acknowledges support from the NVIDIA Academic Hardware 
Grant Program, as well as helpful discussions with Jad Halimeh and Benedikt Kloß regarding disorder-free localisation, and technical discussions with Jan Naumann on the use of the \texttt{JAX} library~\cite{jax2018github} for GPUs, which was used throughout this work, as well as HPC support from Jens Dreger.
All data and code used in this project are available at Refs.~\cite{PyFlow,data}.

\paragraph{Author contributions}
The tensor flow equation 
simulations have been performed by SJT, using the \texttt{PyFlow} package~\cite{PyFlow}. The LIOM norm calculations have been performed by CB, using data from \texttt{PyFlow} processed with \texttt{QuSpin}~\cite{Weinberg+17,Weinberg+19}, in order to build the LIOM operators in matrix form. All authors conceived the project and contributed to the final manuscript.
\newpage

\appendix
\onecolumngrid
\section{Error measures}
\label{sec.checks}

In this Appendix we will briefly review several error estimation methods used to verify the accuracy of our results. 

\begin{figure}[h!]
    \centering
    \includegraphics[width=\linewidth]{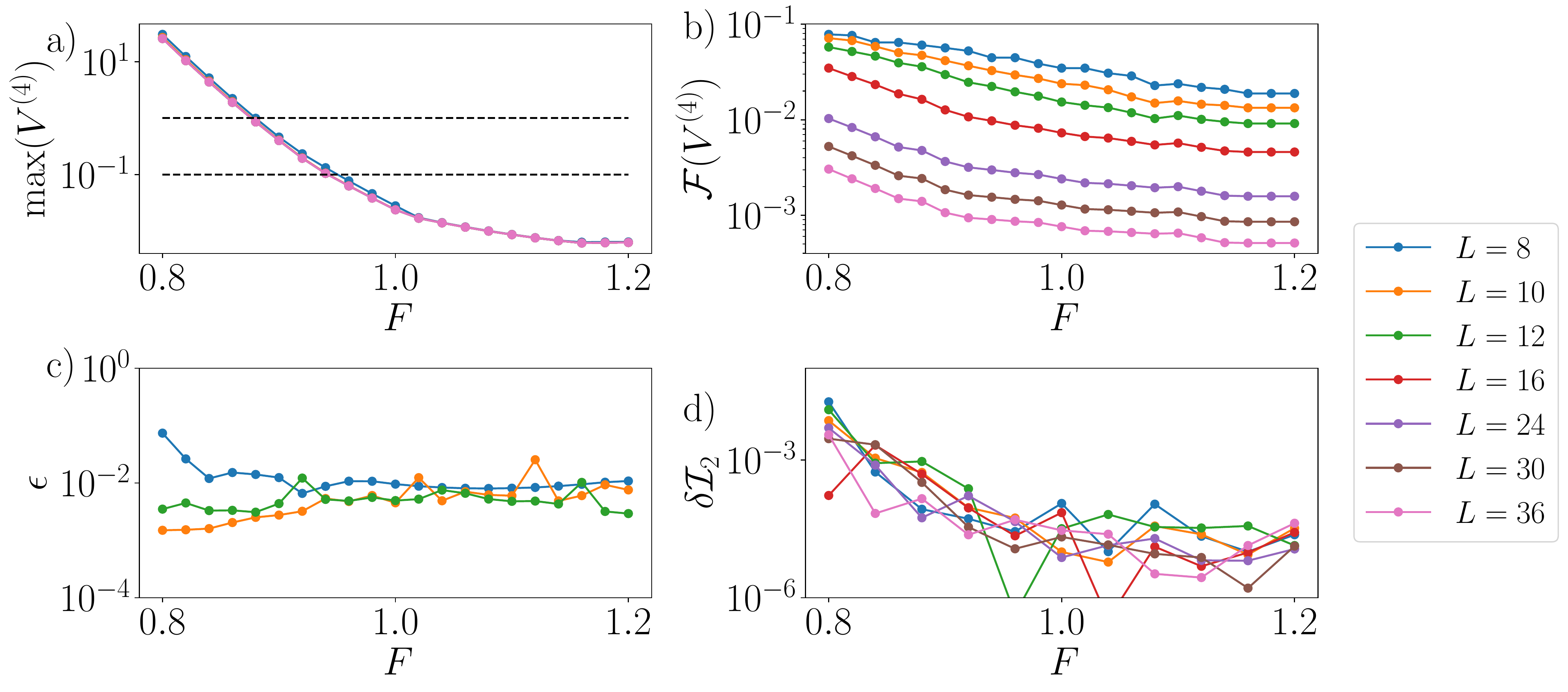}
    \caption{A summary of various error measures used to verify the accuracy of the method, computed for various system sizes as shown in the legend. All panels show results for the linear potential. a) The magnitude of the maximum quartic off-diagonal element at the end of the flow. (Note that all system sizes approximately overlap.) b) The fraction of quartic off-diagonal elements which have not decayed below a cutoff of $\varepsilon = 10^{-3}$ by the end of the flow, confirming that any non-decayed couplings form a vanishingly small fraction in the thermodynamic limit. c) The relative error of the diagonal Hamiltonian computed with respect to exact diagonalisation for small system sizes. d) A self-consistent measure of the deviation from unitarity, measured with respect to the diagonal final Hamiltonian, as detailed in the text. This self-consistent measure is approximately independent of system size.}
    \label{fig.checks}
\end{figure}

\subsection{Convergence}

Flow equation methods based on Wegner-type generators are known to be unable to fully diagonalise systems with degeneracies, as the method requires a clear separation of energy scales in order to transform away off-diagonal elements. As the linear potential here leads to a highly degenerate many-body spectrum, it is important to check carefully the convergence properties of the algorithm. One can easily see that there are no single-particle degeneracies, however,
the interacting part of the Hamiltonian is indeed highly degenerate, and as such some off-diagonal elements of the quartic component of the Hamiltonian can survive until the end of the flow, and can even diverge sufficiently far into the delocalised phase where the method breaks down.

The magnitude of the maximum quartic off-diagonal element remaining at the end of the flow $\textrm{max}(V^{(4)})$ is shown in Fig.~\ref{fig.checks}a). We see that it increases as $F$ becomes small, consistent with our understanding that in the delocalised phase the method is unable to fully diagonalise the Hamiltonian, and that it is almost independent of system size. This measure is a worst-case error estimate, however, as the terms which fail to decay are an extremely small fraction of the total number of off-diagonal terms. We demonstrate this in Fig.~\ref{fig.checks}b) where we plot the fraction of off-diagonal terms remaining at the end of the flow $\mathcal{F}(V^{(4)})$ which are above a cutoff value of $\varepsilon = 10^{-3}$. We see that this quantity is extremely small and decreases with both increasing system size $L$ and increasing slope $F$, confirming that the resonant couplings which fail to decay under the action of the transform constitute a vanishing fraction of the total number of terms in the Hamiltonian. In fact, careful examination reveals that the divergent terms are almost always unphysical, containing two creation/annihilation operators which act on the same site, and are usually confined to the boundaries of the system. These terms cannot, therefore, 
act on any physical state.

In the following error estimates we will neglect these off-diagonal terms, as their inclusion would force us to build the Hamiltonian as an exponentially large matrix in the full Hilbert space, whereas neglecting them allows us to easily and straightforwardly compute various quantities of the (approximately) diagonal final Hamiltonian. It must be understood that the convergence shown in this section is the primary check of the accuracy of the method, and only once convergence is assured can the following methods be used.

\subsection{Comparison with exact diagonalisation}

The next error estimation we can make is a comparison of the eigenvalues obtained from the final diagonal Hamiltonian with those obtained by numerically \emph{exact diagonalisation} (ED), computed using the \texttt{QuSpin} package~\cite{Weinberg+17,Weinberg+19}. Clearly, this comparison is only valid for small system sizes, but it is nonetheless an instructive and useful sanity check before moving on to self-consistent error estimates that can be applied to systems larger than those which can be studied with ED.

In order to compare the results, we simply apply the diagonal Hamiltonian to all $2^L$ basis states and compute their energies, then compare this with the spectrum returned from ED. We define the relative error as
\begin{align}
    \epsilon := \frac{1}{2^L}\sum_n \left| \frac{E_n^{ED}-E^{FE}}{E_n^{ED}}\right| .
\end{align}
This quantity is shown in Fig.~\ref{fig.checks}c), for system sizes $L=8,10$ and $12$ only. We see that the relative error remains on the order of $1\%$ for all values of $F$.

\subsection{Invariants of the flow}

As the previous error estimate is only useful for small systems where exact diagonalisation can be used, we also make use of a second measure, namely the invariants of the flow. This relies on the fact that unitary transforms conserve integer powers of the Hamiltonian, i.e., $I_p(l) = \textrm{Tr}[\mathcal{H}^p(l)]$ (where $p$ is an integer) will be the same for any value of $l$. These are known as invariants of the flow. We define the following quantity which measures the difference between the $p$-th invariant at the start and end of the flow and acts as a measure of how far our approximate transform deviates 
from unitarity,
\begin{align}
    \delta I_p = \left| \frac{I_p(l=0) - I_p(l \to \infty)}{I_p(l=0)} \right|.
\end{align}
Here, we specify to the case of $p=2$, which can be straightforwardly computed if we restrict the trace to the basis containing $m$ particles, where $m \ll L$. We use the \texttt{QuSpin} library to build the Hamiltonian as a matrix in the basis of $m$-particle states (using the output from the TFE method) and compute the trace exactly numerically. The results are shown in Fig.~\ref{fig.checks}d) for all system sizes, where we compare the diagonal final Hamiltonian with the initial Hamiltonian. Here we use $m=2$, the smallest non-trivial number of particles that allows us to probe the error in the interaction term of the Hamiltonian. Similar results are obtained for 
$m \in \{4,5,6\}$, 
however, the memory requirements quickly become unfeasible for $m>4$ at the largest system sizes we consider, and so we specify to $m=2$ here. We see that the flow invariant is well-conserved throughout, confirming that the transform remains close to unitary.

\begin{figure}
    \centering
    \includegraphics[width=0.8\linewidth]{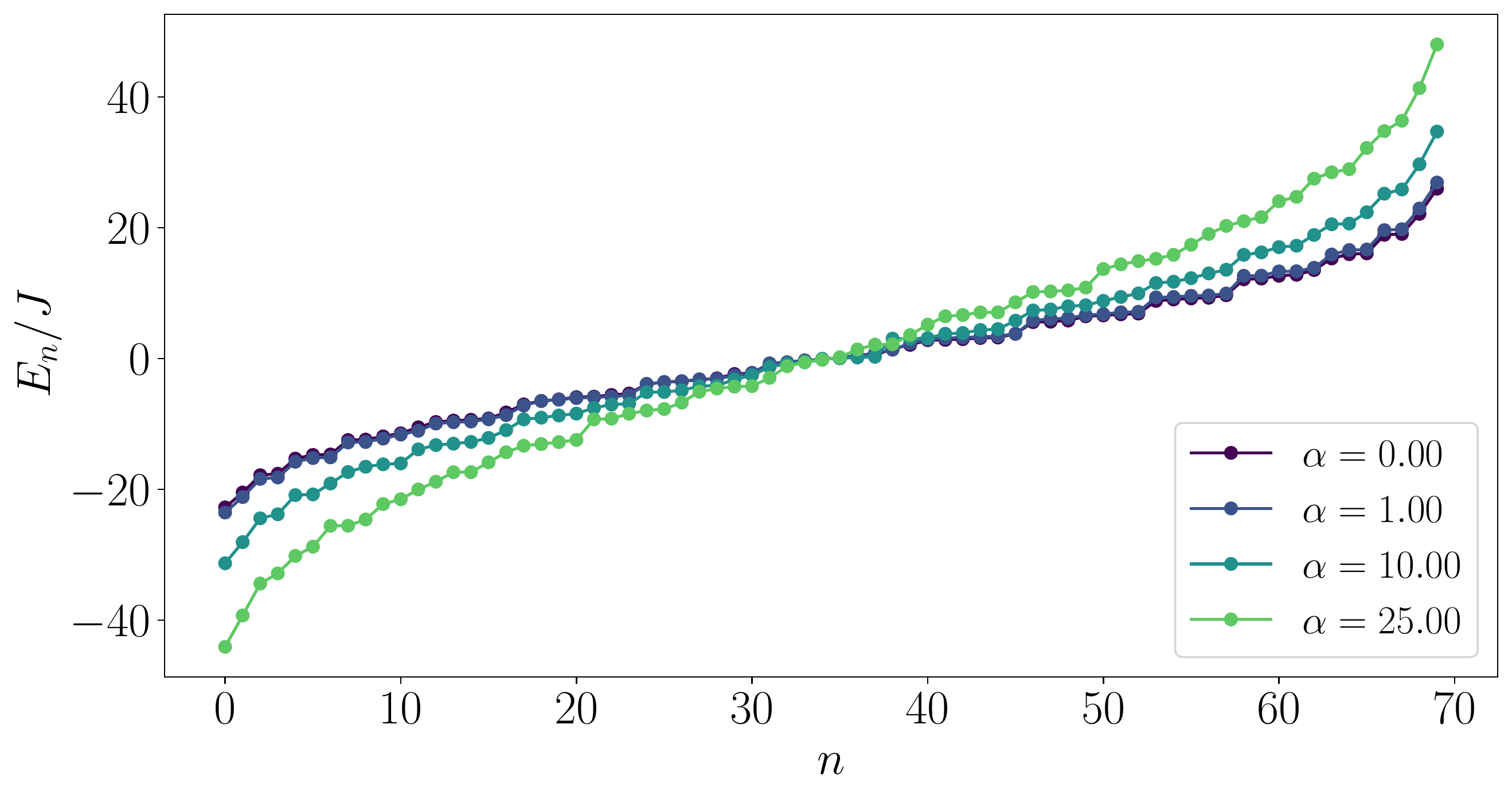}
    \caption{The energy spectrum of a small system of size $L=8$ at half-filling, with $F/J=3.0$ and $\Delta_0=1.0$. Data is shown for four different values of the curvature parameter $\alpha$ to illustrate the degeneracies that exist in the spectrum of the linear potential ($\alpha=0$) and how these degeneracies are lifted by increasing values of $\alpha$. }
    \label{fig.degeneracies}
\end{figure}

\section{Effects of curvature}
\label{app.curvature}

Here we briefly show exact diagonalisation results for the energy spectrum of a small system of size $L=8$ to demonstrate the degeneracies present in the case of a purely linear potential, and how they are lifted in the presence of some curvature. The results are shown in Fig.~\ref{fig.degeneracies}. It is interesting to note that even for very large curvature, some near-degeneracies remain, suggesting that curvature is not able to fully remove the many-body resonances which are thought to lead to delocalisation. For small systems, curvature is able to lift the majority of the degeneracies, but with increasing system size the number of degeneracies increases faster than moderate curvature can compensate for (as we discuss in Section~VI of the main text).

\begin{figure}
    \centering
    \includegraphics[width=\linewidth]{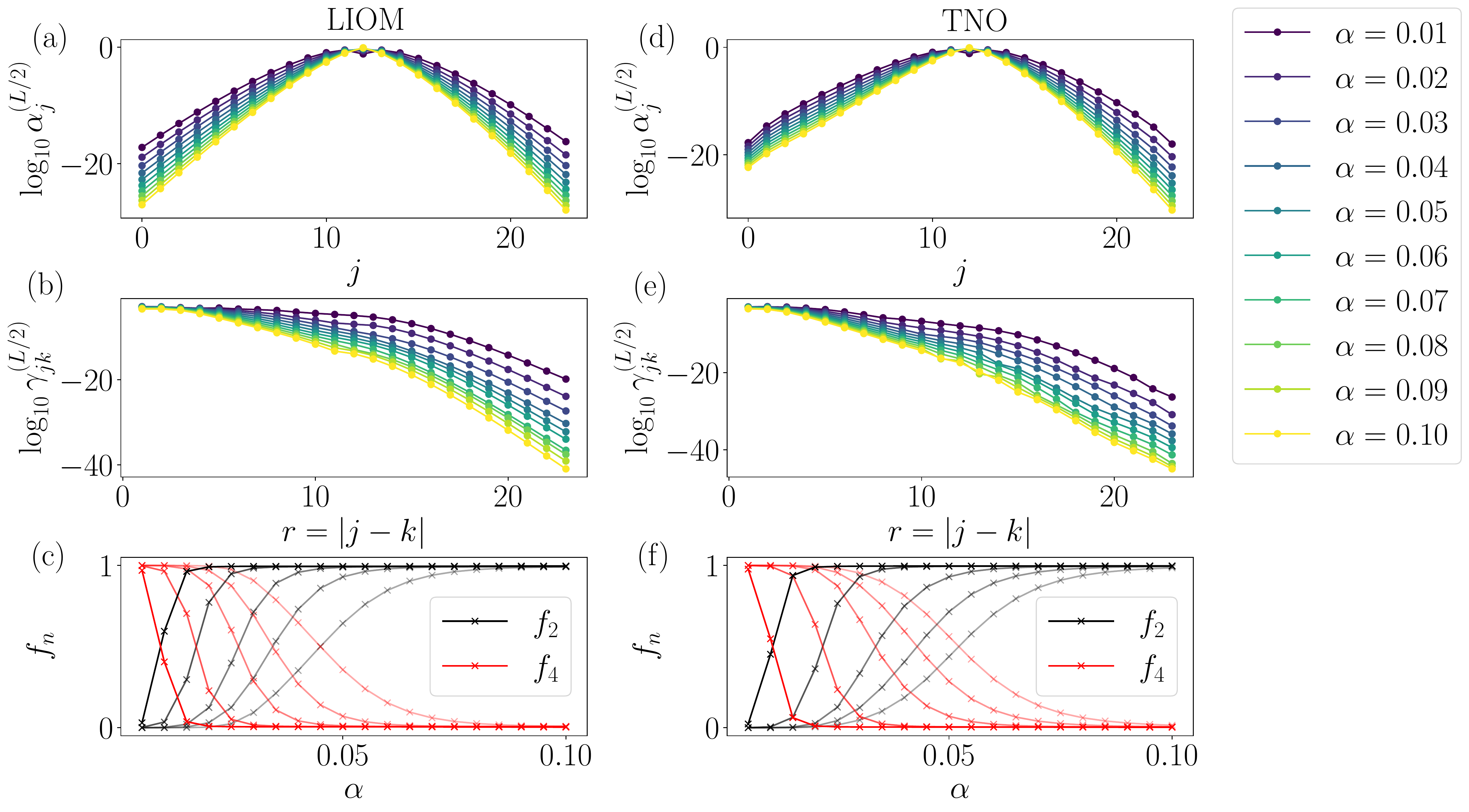}
    \caption{The same quantities as shown in Fig.~9 in the main text, but now computed for the quadratic potential \emph{without} the factor of $1/L^2$. Unless stated, the results are for system size $L=24$ with fixed slope $F=0.8$. a) The quadratic part of the LIOM located on the central site of the chain. b) The diagonal quartic part of the LIOM. c) The ratios $f_2$ and $f_4$ for system sizes $L=8,10,12,16,24$ (light to dark), showing an enhancement of the localised phase with increasing system size. d) The quadratic part of the transformed number operator initially located at the centre of the chain. e) The diagonal quartic part of the transformed number operator. f) The ratios  $f_2$ and $f_4$ computed using the transformed number operator, for the same system sizes as in panel c).}
    \label{fig.curved2}
\end{figure}

\section{Curved potentials with no $1/L^2$ dependence}
\label{sec.curved2}
In the main text we considered a curved potential of the form $h_i = F i + \alpha (i/L)^2$, 
however, in the limit of $L \to \infty$ this clearly reduces back to the linear potential, with its associated highly degenerate many-body spectrum. Here we briefly investigate an alternative definition of a curved potential, given by
\begin{align}
    h_i = F i + \alpha i^2
\end{align}
which does not have an explicit dependence on system size. Note, 
however, that for large systems, this potential has its own problem as the quadratic component will quickly dominate over the linear term, however, the lack of dependence on $L$ makes it of interest for finite-size scaling. The results are shown in Fig.~\ref{fig.curved2}, where we plot the same quantities as shown in the main text for the linear and other curved potential. By comparison with the original form of curved potential which includes the factor of $1/L^2$, we see here that the behaviour with system size is reversed, with \emph{larger} systems appearing more localised. This is due to the curved part of the potential dominating for larger systems, as expected, and as such it cannot really be considered to be Wannier-Stark localisation.

%

\end{document}